\documentclass[longauth]{aa}
\pdfoutput=1
\usepackage[varg]{txfonts}
\usepackage[english]{babel}
\usepackage[utf8]{inputenc}
\usepackage[OT2,T1]{fontenc}
\newcommand\textcyr[1]{{\fontencoding{OT2}\fontfamily{wncyr}\selectfont #1}}
\usepackage{paralist}
\usepackage{graphicx}
\usepackage{longtable}
\usepackage{subfig}
\usepackage{float}
\usepackage{natbib}
\graphicspath{{figures/}}

\begin{document}

\bibpunct{(}{)}{;}{a}{}{,}

\title{The UV and Ly$\alpha$ Luminosity Functions of galaxies and the Star Formation Rate Density at the end of HI reionization from the VIMOS Ultra-Deep Survey (VUDS)
  \thanks{Based on data obtained with the European 
	  Southern Observatory Very Large Telescope, Paranal, Chile, under Large
	  Program 185.A--0791. }}

\author{Y.~Khusanova~(\textcyr{Ya. Khusanova})\inst{\ref{LAM}} 
\and O.~Le F\`evre\inst{\ref{LAM}} 
\and P.~Cassata\inst{\ref{Padova}} 
\and O.~Cucciati\inst{\ref{BolognaObs}} 
\and B.~C.~Lemaux\inst{\ref{Davis}} 
\and L.~A.~M.~Tasca\inst{\ref{LAM}}
\and R.~Thomas\inst{\ref{ESOChile}}
\and B.~Garilli\inst{\ref{Milano}} 
\and V.~Le Brun\inst{\ref{LAM}} 
\and D.~Maccagni\inst{\ref{Milano}} 
\and L.~Pentericci\inst{\ref{RomeObs}} 
\and G.~Zamorani\inst{\ref{BolognaObs}} 
\and R.~Amor\'in\inst{\ref{Serena1},\ref{Serena2}} 
\and S.~Bardelli\inst{\ref{BolognaObs}} 
\and M.~Castellano\inst{\ref{RomeObs}} 
\and L.~P.~Cassar\`a\inst{\ref{Milano}} 
\and A.~Cimatti\inst{\ref{BolognaUni},\ref{FirenzeObs}} 
\and M.~Giavalisco\inst{\ref{UMass}} 
\and N.~P.~Hathi\inst{\ref{STSI}} 
\and O.~Ilbert\inst{\ref{LAM}} 
\and A.~M.~Koekemoer\inst{\ref{STSI}} 
\and F.~Marchi\inst{\ref{RomeObs}} 
\and J.~Pforr\inst{\ref{ESTEC}} 
\and B.~Ribeiro\inst{\ref{LeidenObs}} 
\and D.~Schaerer\inst{\ref{GeneveObs}}
\and L.~Tresse\inst{\ref{Lyon},\ref{LAM}}
\and D.~Vergani\inst{\ref{BolognaObs}} 
\and E.~Zucca\inst{\ref{BolognaObs}}
}
\institute{Aix Marseille Universit\'e, CNRS, LAM (Laboratoire d'Astrophysique de Marseille) UMR 7326, 13388, Marseille, France\label{LAM}
\and University of Padova, Department of Physics and Astronomy Vicolo Osservatorio 3, 35122, Padova, Italy\label{Padova}
\and INAF - Osservatorio di Astrofisica e Scienza dello Spazio di Bologna, via Gobetti 93/3, I-40129, Bologna, Italy\label{BolognaObs}
\and Department of Physics, University of California, Davis, One Shields Ave., Davis, CA 95616, USA\label{Davis}
\and European Southern Observatory, Avenida Alonso de Córdova 3107, Vitacura, 19001, Casilla, Santiago de Chile, Chile\label{ESOChile}
\and INAF--IASF Milano, via Corti 12, 20133, Milano, Italy\label{Milano}
\and INAF, Osservatorio Astronomico di Roma, Monteporzio, Italy\label{RomeObs}
\and Instituto de Investigaci\'on Multidisciplinar en Ciencia y Tecnolog\'ia, Universidad de La Serena, Ra\'ul Bitr\'an 1305, La Serena, Chile\label{Serena1}
\and Departamento de F\'isica y Astronom\'ia, Universidad de La Serena, Norte, Av. Juan Cisternas 1200, La Serena, Chile\label{Serena2}
\and Università di Bologna, Dipartimento di Fisica e Astronomia, Via Gobetti 93/2, I-40129, Bologna, Italy\label{BolognaUni}
\and INAF - Osservatorio Astrofisico di Arcetri, Largo E. Fermi 5, I-50125, Firenze, Italy\label{FirenzeObs}
\and Astronomy Department, University of Massachusetts, Amherst, MA01003, USA\label{UMass}
\and Space Telescope Science Institute, 3700 San Martin Drive, Baltimore, MD, 21218, USA\label{STSI}
\and ESA/ESTEC SCI-S, Keplerlaan 1, 2201 AZ, Noordwijk, The Netherlands\label{ESTEC}
\and Leiden Observatory, Leiden University, PO Box 9513, 2300 RA, Leiden, The Netherlands\label{LeidenObs}
\and Observatoire de Gen\`eve, Universit\'e de Gen\`eve, 51 Ch. des Maillettes, 1290 Versoix, Switzerland\label{GeneveObs}
\and Univ Lyon, Univ Lyon1, Ens de Lyon, CNRS, Centre de Recherche Astrophysique de Lyon UMR5574, F-69230, Saint-Genis-Laval, France\label{Lyon}
}

\abstract{
}{
We establish a robust statistical description of the star-forming galaxy population at the end of cosmic HI reionization ($5.0\le{}z\le6.6$) from a large sample of 52 galaxies with spectroscopically confirmed redshifts. Rest-frame UV and Ly$\alpha$ luminosities are used to construct luminosity functions. We calculate star formation rate densities (SFRD) at the median redshift of our sample z=5.6.}
{
We use the VIMOS UltraDeep Survey to select a sample of galaxies at $5.0\le z_{spec}\le 6.6$. We clean our sample from low redshift interlopers using ancillary photometric data. We identify galaxies with Ly$\alpha$ either in absorption or in emission, at variance with most spectroscopic samples in the literature where Ly$\alpha$ emitters dominate. We use the 1/V$_{max}$ method to determine luminosity functions.}
{
The galaxies in this redshift range exhibit a large range in their properties. A fraction of our sample shows strong Ly$\alpha$ emission, while another fraction shows Ly$\alpha$ in absorption. UV-continuum slopes vary with luminosity, with a large dispersion. We find that star-forming galaxies at these redshifts are distributed along a main sequence in the stellar mass vs. SFR plane, described with a slope $\alpha=0.85\pm0.05$ and a dispersion of 0.13 dex. We report a flat evolution of the sSFR(z) in 3<z<6 compared to lower redshift measurements. We find that the UV luminosity function is best reproduced by a double power law with parameters: $\Phi^*=2.5\times10^{-4}$ mag$^{-1}$Mpc$^{-3}$ , $M^*=-21.43_{-0.10}^{+0.13}$, $\alpha=-2.0$, $\beta=-4.52_{-0.48 }^{+0.49}$, while a fit with a Schechter function is only marginally worse. The best fit parameters for the Ly$\alpha$ luminosity function are $\alpha=-1.69$, $\log\Phi^*$(Mpc$^{-3}$)$=-3.21_{-0.10}^{+0.12}$  and $\log{L^*}$(erg s$^{-1}$)$=43.00_{-0.12}^{+0.09}$ for a Schechter function parameterization. We derive a $\log{}SFRD_{UV}$(M$_{\odot}$yr$^{-1}$Mpc$^{-3}$)=$-1.34_{-0.08}^{+0.06}$ and $\log{}SFRD_{Ly\alpha}$(M$_{\odot}$yr$^{-1}$Mpc$^{-3}$)=$-2.02_{-0.08}^{+0.07}$. After we correct for IGM absorption, with the assumption of a low dust content, we find that the SFRD derived from the Ly$\alpha$ luminosity function is in excellent agreement with the UV-derived SFRD. 
}
{Our new SFRD measurements at a mean redshift z=5.6 are 0.2-0.3 dex above the mean SFRD reported in \citet{madau_cosmic_2014}, but in excellent agreement with results from \citet{bouwens_uv_2015} and confirm the steep decline of the SFRD at z>2. The bright end of the Ly$\alpha$ luminosity function has a high number density, indicating a significant star formation activity concentrated in the brightest Ly$\alpha$ emitters (LAE) at these redshifts. LAE with EW>25\AA ~contribute to about 75\% of the total UV-derived SFRD. While our analysis favors a low dust content in 5.0<z<6.6, uncertainties on the dust extinction correction and associated degeneracies in spectral fitting will remain an issue to estimate the total SFRD until future survey extending spectroscopy to the NIR rest-frame spectral domain, e.g. with JWST.
}

  \keywords{
Galaxies: high redshift --
Galaxies: evolution --
Galaxies: formation --
Galaxies: star formation --
Galaxies: luminosity function -- 
Cosmology: reionization
               }

\titlerunning{VUDS: UV and Ly$\alpha$ Luminosity Functions, Star Formation Rate Density at the end of HI reionization}
\authorrunning{Yana Khusanova et al.}

\maketitle

\section{Introduction}

As the first galaxies form, they ionize the local medium they are embedded in, letting radiation free to propagate \citep[e.g.][]{Dayal18}. There is a building consensus that the end of the Hydrogen reionization epoch is at a redshift z$\sim$6, stemming from several lines of evidence. The optical depth of HI reionization is encoded in the standard $\Lambda$CDM cosmological world model, and it can be extracted from observations of the cosmic microwave background (CMB), giving access to the start, mid-point and end of the reionization process \citep{hinshaw_nine-year_2013}. From the WMAP CMB observations, it was inferred that the reionization would have been $\sim$50\% completed at z$\sim$10.5 \citep{spergel_first-year_2003}. This would require a presence of a substantial ionizing background at quite early times, which should have materialized from star-forming galaxies at very early epochs. Early attempts to reconcile WMAP measurements with UV background estimates found it hard to identify enough galaxies capable to produce the required number of UV photons  at early times \citep{robertson_new_2013}. Several hypotheses were proposed in order to solve this discrepancy, among them invoking a substantial population of growing supermassive black holes and numerous faint galaxies, escaping detection \citep{ciardi_early_2003, volonteri_relative_2009}. 

The more recent findings on the epoch of HI reionization from the Planck experiment revisited the issue from the CMB point of view, and significantly reduced the optical depth of reionization,
which lowers the requirements on the number of ionizing photons and hence on the number of galaxies at high redshifts. Comparing the optical depth of reionization from Planck and from deep surveys, \cite{robertson_cosmic_2015} and \cite{bouwens_reionization_2015} claim that galaxies produce enough ionizing photons, provided that there are enough faint galaxies populating the faint-end slope of the UV luminosity function. The latest results from the Planck experiment favor an even smaller reionization optical depth. The redshifts of the start, 50\%, and end of reionization, derived from the CMB Planck maps with 95\%~CL, are now $z=10.4\pm1.8$, $8.5\pm0.9$, $<8.9$ , respectively \citep{planck_collaboration_planck_2016}. These results  further consolidate the picture of  reionization  that  happened  late  and  fast,  and  reionization  being  driven  by  photons  from  massive  stars in low mass galaxies as outlined in the 2018 Planck satellite results \citep{Planck18}.

However, matching the CMB results with galaxy counts remains a considerable challenge. Deep galaxy surveys are constantly pushing the search for galaxies capable to produce the needed ionizing photons, to higher redshifts and fainter luminosities \citep[e.g.][]{LeFevre2005, scoville_cosmos_2007,Stark2009,Grogin2011,Pentericci2011,LeFevre2013,Ellis2013,bowler_galaxy_2015,lefevre_vimos_2015,bouwens_reionization_2015}. The challenge is to characterize the luminosity function of these first galaxies with enough accuracy that the total number of ionizing photons can be accurately estimated. The census of high redshift galaxies is continuously improving, first and foremost on the basis of deep imaging with the Hubble Space Telescope (HST) and selected ground-based facilities. Faint multi-band photometry reaching magnitudes AB$\sim$30 significantly increased the number of galaxy candidates with z$>$6 and up to z$\sim$10, from the HST CANDELS \citep{Grogin2011, koekemoer_candels:_2011}, COSMOS \citep{scoville_cosmos_2007}, Frontier Fields \citep{Finkelstein2015, lotz_frontier_2017} and UltraVista surveys \citep{McCracken2012}. The additional boost from gravitational lensing allows to further constrain the faint-end slope of the luminosity function, reaching M$_{UV}\sim-13$ \citep{livermore_directly_2017, bouwens_z6_2017, ishigaki_full-data_2018, Yue2018}. These surveys form the basis of our understanding of the UV rest-frame luminosity function and the derived Star Formation Rate Density \citep[SFRD,][]{madau_cosmic_2014}, at these redshifts. However, these observations remain difficult, and improving the faint galaxy census at z>5 from high purity and completeness counts of galaxies with confirmed redshifts therefore remains of the utmost importance, particularly to set robust constraints on the SFRD history.

In this paper we focus on providing robust counts of galaxies covering a redshift range from z$\sim$5 to z$\sim$6.6, a time close to, or including, the end of reionization. This corresponds to a cosmic time period from 0.8 to 1.15 Gyr after the Big Bang. We base our counts on a sample of galaxies with a reliable {\it spectroscopic redshift} identification obtained from the VIMOS Ultra-Deep Survey  \citep[VUDS,][]{lefevre_vimos_2015}, at variance from most previous studies based on photometric redshift identification \citep{mclure_luminosity_2009,bouwens_uv_2015,bowler_galaxy_2015} or spectroscopy with narrower selection criteria \citep{ouchi_subaru/xmm-newton_2008,cassata_vimos_2011,santos_lyman-alpha_2016, drake_muse_2017}. We identify and characterize star-forming galaxies, focusing on several key quantities, including the UV rest-frame luminosity, the Ly$\alpha$ line luminosity, as well as stellar mass, star formation rates (SFR), and other parameters derived from spectral energy distribution (SED) fitting. We derive cosmic SFRDs from observed rest-frame UV luminosity functions as well as from Ly$\alpha$ luminosity functions.

The paper is organized as follows. In Sect. \ref{data} we describe our methods to isolate a reliable sample of 52 galaxies with spectroscopic redshifts $5<z<6.6$.
We present the sample in Sect. \ref{sample}, including the redshift distribution, average spectral properties, UV $\beta$-slopes, and the distribution of these galaxies in the SFR versus stellar mass diagram. Observed galaxy counts are used with UV and Ly$\alpha$ luminosities to derive luminosity functions in Sect. \ref{LFs}. We then derive SFRDs from the UV and Ly$\alpha$ luminosity functions, compare them and discuss their evolution with redshift in Sect. \ref{sfrd}. The results are summarized in Sect. \ref{summary}.

Throughout the paper we use $\Lambda$CDM cosmology with $H_0=70$ km/s/Mpc, $\Omega_{\Lambda}=0.70$, $\Omega_m=0.30$. All magnitudes are given in the AB system.

\section{Data}
\label{data}

\subsection{Spectroscopic and photometric data}
\label{spec_phot}

We use the  spectroscopic sample of galaxies drawn from the VUDS, which is described in detail in \cite{lefevre_vimos_2015}. The wavelength coverage of the survey is from 3650 to 9350 \AA~and enables secure redshift measurements up to redshift z$\sim$6.6, when the Lyman-$\alpha$-1215\AA~line leaves the spectral window. The spectra allow to follow important spectral features to guarantee unambiguous spectroscopic redshifts accurate to $\sim$$10^{-3}$ \citep{lefevre_vimos_2015}. Most of the spectra in the survey were observed with a low resolution grating  ($R=230$). Complementary to it, a number of objects were observed with a medium resolution of $R=580$, observed in priority in the MOS masks for having $z_{phot}>4.5$.

The survey covers three different fields in VVDS02h, COSMOS and ECDFS for a total area of 1 deg$^2$, minimizing the effects of cosmic variance. A wide range of ancillary data is available for each field to produce a sample for which completeness and purity are well controlled, and to infer physical parameters, most importantly stellar masses and SFRs through SED fitting performed with the knowledge of the accurate redshift.

In the VVDS02h field we use photometry from the 7th data release of CFHT Legacy Survey \citep[CFHTLS, ][]{Cuillandre2012}, which covers the $u^*, g', r', i', z'$ optical bands. It is complemented by infrared data in $J, H, K$ bands from WIRCam Deep Survey \citep[WIRDS, ][]{Bielby2012} and in two IRAC bands (3.6 and 4.5 $\mu$m) from the Spitzer Extragalactic Representative Volume Survey \citep[SERVS, ][]{Mauduit2012}.

The full range of photometric observations in the COSMOS field is presented on the COSMOS web site\footnote{http://cosmos.astro.caltech.edu/}. In this work, we use the optical data from CFHT for $u^*$ band and Subaru broad bands $B, V, g+, r+, i+, z+$, the near-infrared bands $J$ and $K$ from UltraVista \citep{McCracken2012} and IRAC bands from the Spitzer Extended Deep Survey \citep[SEDS, ][]{Ashby2013}.

In the ECDFS field we use, depending on availability, either the CANDELS set of photometry \citep{Grogin2011, koekemoer_candels:_2011}, which includes observations in the optical and near-infrared HST bands, or the Taiwan ECDFS Near-Infrared Survey data (TENIS) with observations $J$ and $K_s$ bands \citep{Hsieh2012}, complemented by the Galaxy Evolution from Morphology and SEDs survey (GEMS) in F606W and F850LP bands\citep{Caldwell2008}.

The main target selection is based on photometric redshifts and is designed to go to the highest possible redshifts starting from $z=2$. The galaxies were chosen to have $z_{phot} + 1\sigma \ge 2.4$, where $z_{phot}$ is either the first or the second peak in the photometric redshift probability distribution function (PDF). The limiting magnitude of the main survey target selection is $i=25$. This primary sample was complemented by galaxies chosen with two widely used techniques. The first one is the Lyman-break or dropout technique, which is based on the search of a break in the continuum corresponding to the changing spectral shape of a galaxy continuum between the Lyman limit at 912\AA~and Ly$\alpha$, classically followed in color-color diagrams \citep{Steidel1996}. The second technique is based on the search of Ly$\alpha$ line emission in narrow bands. The galaxies from both these selections may have $i>25$. The galaxies chosen with the dropout technique have limiting magnitude $K_{AB}\le24$, and are added to the target sample only if they are not already selected in the primary sample. In our high redshift sample we have about 10\% of the sample chosen by these criteria.

In addition to this main sample, in the process of examining all 2D spectra visually, the VUDS team discovered a number of single emission lines, belonging to objects falling by chance in the slits, but for which no counterparts could be identified on any of the available images. These were analyzed following a method similar to that of \cite{cassata_vimos_2011} to assess the nature of the line. In this way, we identified a number of serendipitous Ly$\alpha$ emitters (LAE) with a UV continuum flux too faint to be detected in broad photometric bands.

The combination of these different samples with complementary selection functions results in a well-defined sample  identified  at $z>5$, covering a broad range of properties. The selection of each sub-sample needs to be fully taken into account in determining luminosity functions, because the galaxies chosen by different techniques are drawn from different parent populations, as discussed in Sect. \ref{LFs}.

Spectroscopic redshifts were measured using the EZ tool \citep{Garilli2010}. The redshift reliability flags adopted are described in \cite{lefevre_vimos_2015} and correspond to the following probabilities to be correct: flag 1: 50-75\%; flag 2: 75-85\%; flag 3: 95-100\%; flag 9: $\sim80$\%. Spectroscopic redshift measurements using the observed spectral range $3650-9350$\AA~are challenging at $z>5$ due to not only low continuum fluxes but also to the variable noise added by the Earth atmosphere for ground-based observations. Even with 14h integrations with VIMOS on the 8m ESO-VLT, a high fraction of galaxies still has insecure redshifts with reliability flags 1 at these redshifts. Another challenge is the possible confusion between the Ly$\alpha$ and the [OII] emission lines, due to the resolution of our spectra. We cannot resolve the [OII] doublet or an asymmetric structure of Ly$\alpha$ line on the spectra of individual objects at the observed spectral resolution. Therefore additional verification of the measured redshifts is needed. We scrutinize each of the $z>5$ candidates following a clear reference protocol to further assess the reliability of their redshifts, and describe the procedure of cleaning the sample from low redshift interlopers in Sect. \ref{candidateselection}.

Multi-band deep imaging helps to further clean the $z>5$ sample from low redshift interlopers. With direct image examination we identify cases in which spectra belongs to two close companions at different redshifts. We then attempt to disentangle true high redshift object from low redshift interloper, combining the spatial location of spectral features observed on the 2D spectrograms with the location of objects observed in images in different wavebands (e.g. indicating a possible continuum dropout signaling a Lyman break for one of the objects).

We also use the photometric measurements to perform SED fitting of each candidate, first without fixing the redshift. This SED fitting provides the PDF of redshifts and the best fit template at the photometric redshift, taken as the peak of the PDF. In order for the SED fitting to be helpful in distinguishing between true high redshift galaxies and low redshift interlopers as described in Sect. \ref{candidateselection}, we need to use a wide range of templates, suitable for both high and low redshift. We use LePhare \citep{Arnouts2002, Ilbert2006} to fit the SED and \cite{bruzual_stellar_2003} models with \cite{Chabrier2003} initial mass function, two star formation histories -- due not only to exponentially declining and delayed and solar and sub-solar metallicities (Z=0.02 and Z=0.008). We consider two extinction laws: \cite{Calzetti2000} and SMC-like \citep{Prevot1984} with E(B-V) in the range from 0.0 to 0.5. The best fit is determined by minimization of $\chi^2$.

The SED fitting, using the spectroscopic redshift, delivers us the best estimate of physical parameters and the best fit template at the spectroscopic redshift. We define the uncertainty on physical parameters from the probability distribution of each parameters. For the FUV magnitudes we use photometric errors to get a robust estimate of uncertainties  (see Sect \ref{UVLF_section}). In Sect. \ref{sample} we extent up to z=6.6 the work of \cite{tasca_evolving_2015}, which was done up to z=5.5. When comparing different physical parameters and their evolution with redshift, we therefore use consistent methods.

\subsection{Candidate selection}
\label{candidateselection}

We start with selecting all galaxies from VUDS with spectroscopic redshifts in the range of our interest 5.0<z<6.6. We select 111 candidates with all reliability flags. Seven of these candidates have the most secure redshifts with flags 3 or 4. In most of the spectra the observed features are either one emission line, which was associated to Ly$\alpha$ in the redshift measurement process, or a continuum break, which was associated to the break at 1215\AA~produced by the strong intergalactic medium (IGM) extinction at these redshifts \citep[e.g.][]{thomas_vimos_2017}. Up to redshift 5 we can usually distinguish between Ly$\alpha$ and [OII] emission lines, because H$\beta$ and the [OIII]-4959/5007\AA~doublet would still observed in the VIMOS spectral window if  a single line with $\lambda<7200\AA$ ~was [OII]-3727\AA ~rather than Ly$\alpha$. At the higher redshifts considered in this paper, H$\beta$ and [OIII] would be shifted beyond the observed wavelength range, and therefore the detection of a single emission line with $\lambda>7200\AA$ ~should be interpreted with caution as discussed  below. Another degeneracy in redshift measurement comes from the possibility to misinterpret a Balmer break at $\sim$$4000$\AA~in a spectrum as a Lyman break or dropout in the continuum due to neutral gas absorption along the line of sight. These degeneracies impose that each high redshift galaxy candidate has to be scrutinized to identify possible low redshift galaxies contaminating our sample. To solve this issue, we make use of all the spectroscopic and ancillary photometric and imaging data and inspect each galaxy individually, imposing the following criteria for a galaxy to be retained in the final $5.0<z_{spec}<6.6$ sample:

\begin{asparaitem}
\item No detection in photometric bands corresponding to wavelength below the Lyman limit at 912\AA, as neutral gas in the galaxy itself should not let photons
out;
\item No detection or weak detection in photometric bands corresponding to wavelengths $912\AA<\lambda<1215.7\AA$, as neutral gas in the intervening IGM should absorb most
of the photons emitted by the galaxy \citep[depending on the line of sight;][]{thomas_vimos_2017};
\item At least one detection in a photometric band corresponding to the rest frame 1215.7\AA ~(for LAE), or at wavelengths longer than 1215.7\AA ~for galaxies without Ly$\alpha$ in emission, as some continuum photons should be detected;
\item The difference between spectroscopic and photometric redshifts should be $z_{spec}-z_{phot}\le3\sigma$, with $\sigma$ being the halfwidth of the 68.3\% confidence interval around the peak in the photometric redshift PDF, for at least one of the peaks in the photometric redshifts PDF;
\item The best SED fit template with the redshift fixed at the spectroscopic redshift and based on photometric points is in agreement with the observed spectrum;
\end{asparaitem}

Galaxies with redshift reliability flags 1 and 9 will be affected the most by the degeneracies described above, as for flag 1 spectra are generally of low S/N on the faintest galaxies, and for flag 9 only a single  feature was identified in the spectrum. To be retained in our final sample, we therefore require these galaxies to pass all the above criteria. The galaxies with reliability flags 2-4 have a higher probability to be correct and therefore we exclude them from the final sample only if they do not pass more than one of the above criteria.

This procedure is motivated by the fact that various effects can affect the photometry of the galaxy and the spectroscopy of each galaxy should remain the primary source of information. We pay special attention to the following cases:

\begin{asparaitem}
\item The PDF is very wide and therefore it is not possible to have a robust estimate of the photometric redshift. In this case, we trust the spectroscopic redshift (this is the case for 6 galaxies in the sample);
\item The sky subtraction is less reliable in the presence of bright nearby objects. This can lead to a higher uncertainty on the measured photometric magnitudes for the faint objects and therefore the photometry can be misleading;
\item The photometry indicates a foreground object along the line of sight, or a blend with a nearby object, or is affected by artifacts from the image processing;
\item We find evidence that the observed object is an AGN (e.g. has a broad Ly$\alpha$ emission line) and therefore can be variable;
\end{asparaitem}

Together with the spectra (1D and 2D), photometry and SED of each galaxy, we analyze all images and look for the evidence of such effects. If the 2D spectrum and the images are consistent with the galaxy being at high redshift, but some evidence for contamination is found, we keep it in our sample, but we note that the photometry of this galaxy should be used with caution. Such an example is shown in Fig. \ref{contaminated} where the high redshift object is close to a foreground object affecting both the photometry and spectrum. 

In Table \ref{table:criteria} we present summary of the selection criteria for each galaxy. We find 6 galaxies, which have $z_{spec}-z_{phot}>3\sigma$. 3 of them show signs of contamination by nearby objects and 2 have very narrow peaks in PDF, which are $3\sigma$ to $6\sigma$ away from the spectroscopic redshift but the best fit template to the photometric redshift cannot explain the observed spectrum. We, therefore, keep these galaxies with their spectroscopic redshift. We also keep two galaxies with faint detections below 912\AA, since they have reliable spectroscopic redshifts and SED fitting clearly points to the same redshift, as spectroscopy, while the detections below 912\AA~are likely to be spurious.

The criteria above help us to efficiently get rid of low redshift interlopers as well as to analyze the reliability of the photometry, which we use later to derive FUV-fluxes and physical properties.

Other possible contaminants of our sample include late-type stars, like late M-types or brown dwarfs. When we measure redshifts with EZ  we use a library of star and galaxy templates to fit the observed spectrum by chi-square minimization algorithm. If we only observe continuum in spectra without emission lines, we compare best fit with a star template with the best fit from galaxy templates and if no star template can reproduce the observed spectrum better than the galaxy template, we save the object in our sample, otherwise we conclude, that we observe a star. 

During the inspection of spectra, we find one quasar with a broad Ly$\alpha$ line (FWHM $\sim4300$ km/s) at a redshift z=5.472. We use the luminosity function of \cite{mcgreer_faint_2018} to estimate the probability of finding  quasars in our sample. We integrate the luminosity function down to $M_{FUV}=-21.4$ (the range of absolute magnitudes corresponding to completeness limits of the parent catalogue of our sample) and we multiply it with the cosmic volume of the survey. Assuming a Poisson distribution, we find the probability of finding more than one quasar to be less than 0.7\%. Therefore in these range of absolute magnitudes we expect to have a clean sample of only star forming galaxies, after exclusion of this one quasar discovered in the sample.

\begin{figure}[h!]
  \resizebox{\hsize}{!}{\includegraphics{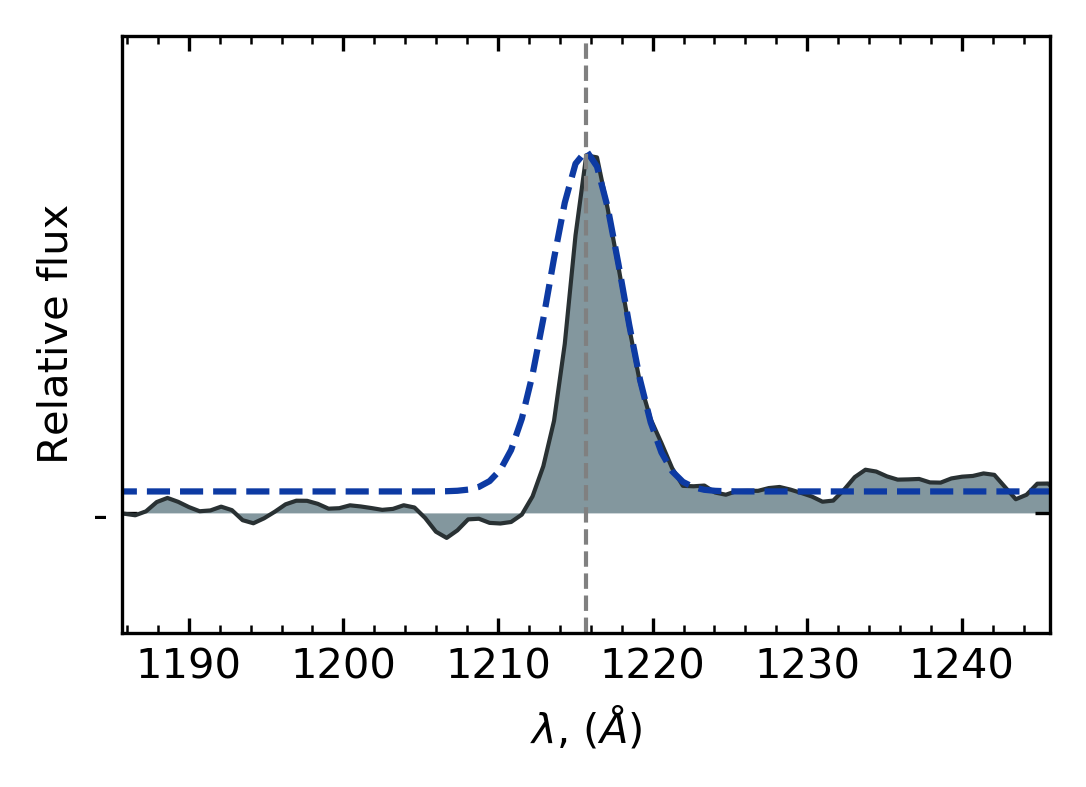}}
  \caption{Stacked spectrum of 36 galaxies with Ly$\alpha$ in emission. The solid line and shaded area below is the observed spectrum, the dashed blue line is a gaussian, fitted to the red wing of the emission line. The vertical dashed line indicates the position of  Ly$\alpha$ line $\lambda=1215.7\AA$.}
  \label{lya_stack}
\end{figure}

We cannot apply the above described criteria to the galaxies with a single emission line and without a photometric counterpart or with contaminated photometry. We therefore need a different way of investigating the reliability of their redshifts. One of the arguments in support of observing Ly$\alpha$ is the skewness of the emission line. The $Ly\alpha$ line has an asymmetric shape with a positive skewness, due to radiation transfer effects, while the unresolved [OII] doublet is usually symmetric with a skewness close to zero. Another effect at high redshift comes from the IGM and circumgalactic medium (CGM), which absorb the continuum below Ly$\alpha$. Due to this effect the $Ly\alpha$ line becomes even more asymmetric.

These effects are difficult to observe on single low signal-to-noise spectra of individual objects at the observed spectral resolution, but the structure of the emission line can be estimated on stacked spectra. We show in Fig. \ref{lya_stack} the stacked spectrum around Ly$\alpha$ of LAE from our sample. We clearly see that the continuum at wavelengths bluer than Ly$\alpha$ is completely absorbed by IGM and CGM and the shape of the line is asymmetric. The measured skewness of the line is SK=2.05, a value higher than reported by \cite{cassata_vimos_2011} SK=1.73 for redshifts $4.6<z<5.9$, but consistent with expectations that at the higher redshifts of our sample, the IGM absorption would be higher. \cite{cassata_vimos_2011} also report a value SK=2.02 for $5.9<z<6.6$ (measured from only 6 galaxies), comparable with our result within uncertainties. The fact that the stacked spectrum of our galaxies has a very strong skewness gives further confidence that our sample is free from the contamination of low redshift objects.
\section{Final sample of 5.0<z<6.6 galaxies}
\label{sample}

\begin{figure}[h]
  \resizebox{\hsize}{!}{\includegraphics{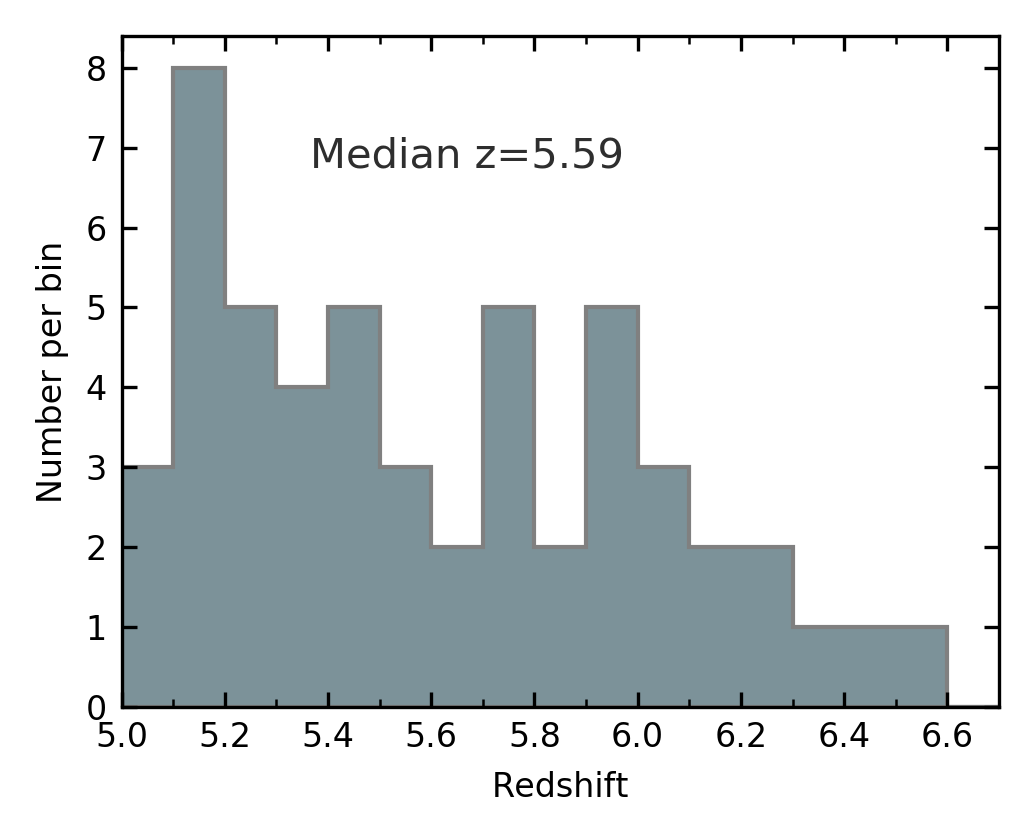}}
  \caption{Redshift distribution of our sample.}
  \label{z_dist}
\end{figure}

After the procedure described above we obtain a sample of 49 galaxies with secure spectroscopic redshifts, 8 of them observed with a medium resolution. In addition, we include 3 galaxies with low signal to noise ratio (SNR) of the emission line (SNR<5.0), which are likely to be at high redshift, but due to the lack of photometric data or spectroscopy at longer wavelength, we cannot confirm their redshift. Since these galaxies are faint both in the Ly$\alpha$ and in the FUV, they do not affect our estimations of the bright end of luminosity functions, and we keep them in our sample to maintain a high completeness. The redshift distribution of the sample is shown in Fig. \ref{z_dist}. The median redshift is $z=5.59$. For 32 of them we derive physical properties and FUV magnitudes from ancillary photometric data (the wavelength coverage does not allow us to measure FUV flux from spectra).

In the subsections below we describe the average and individual properties of the galaxies in our sample.

\subsection{Average spectral properties}

\begin{figure}[h]
  \resizebox{\hsize}{!}{\includegraphics{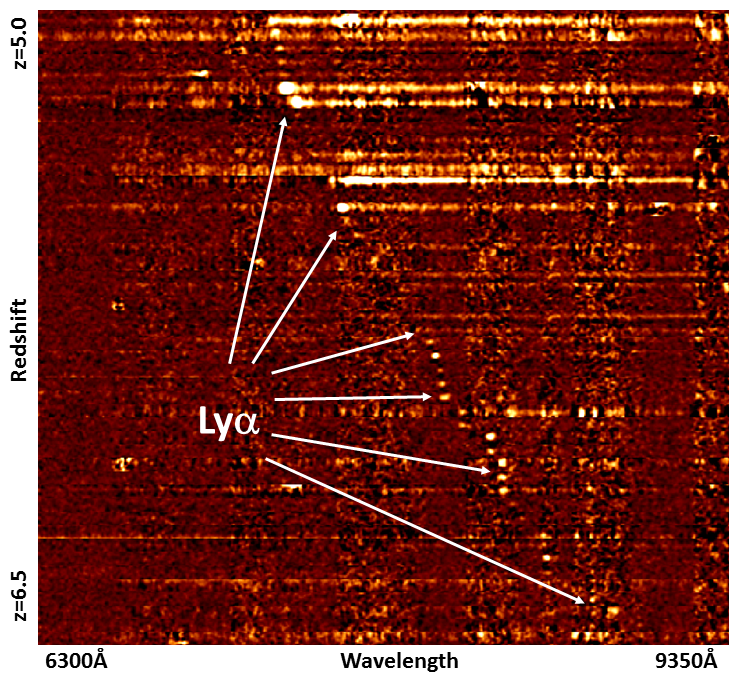}}
  \caption{2D spectra of VUDS galaxies with $5.0\leq z \leq 6.6$, ordered by redshift. The spectral range covers from 6300 to 9350\AA~following the spectral region around Ly$\alpha$. Ly$\alpha$ appears in emission or in absorption, depending on spectra. }
  \label{ordered_spectra}
\end{figure}

The 1D spectra of the individual galaxies together with the best fit SED template and images can be found in the  Appendix. We present the 2D spectra ordered by redshift in Fig. \ref{ordered_spectra}.

\begin{figure}[h]
  \resizebox{\hsize}{!}{\includegraphics{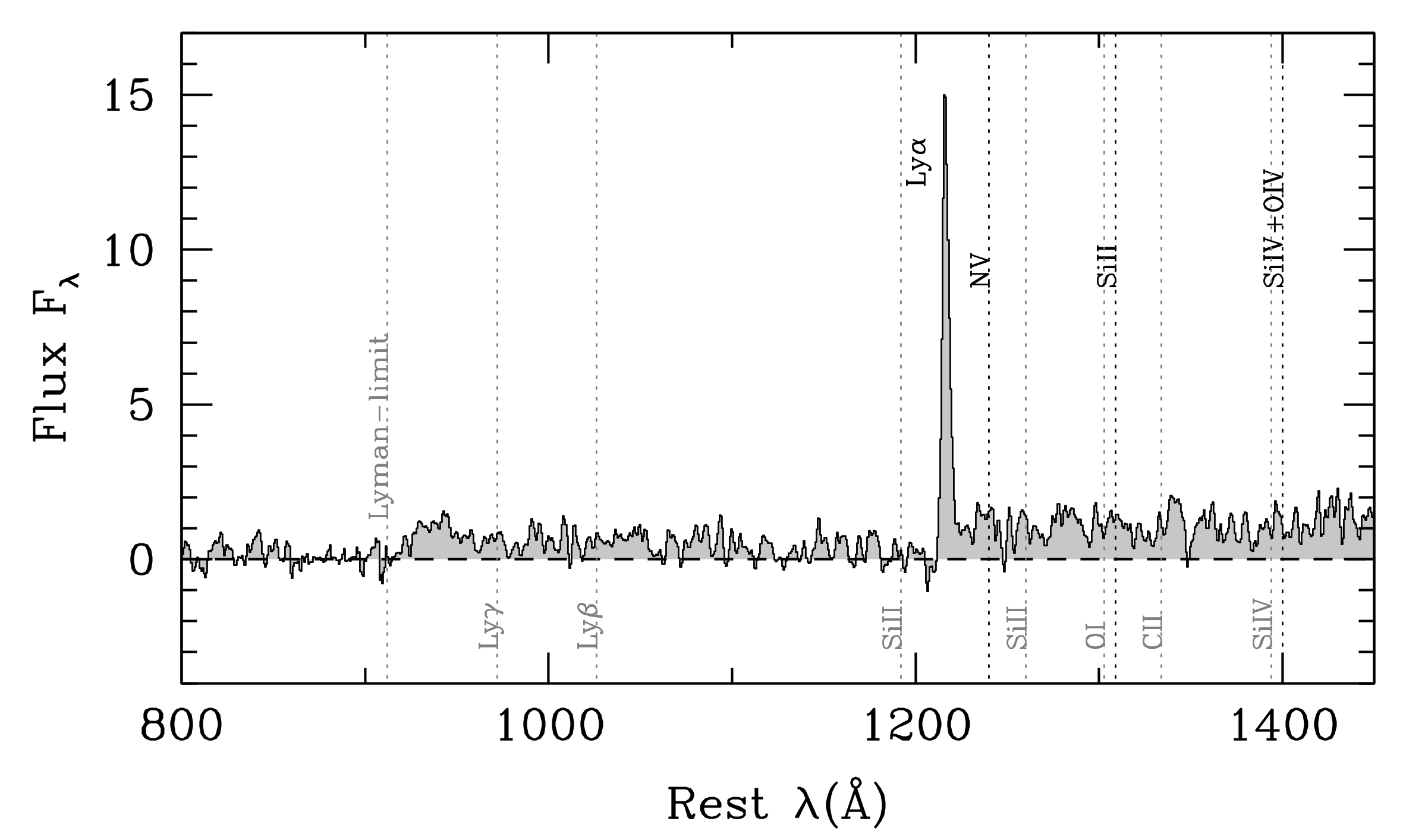}}
  \caption{Median stack of 36 spectra at $5.0 \leq z \leq6.6$ (median $z\sim5.6$)  with Ly$\alpha$ in emission.}
  \label{spec_em}
\end{figure}

\begin{figure}[h]
  \resizebox{\hsize}{!}{\includegraphics{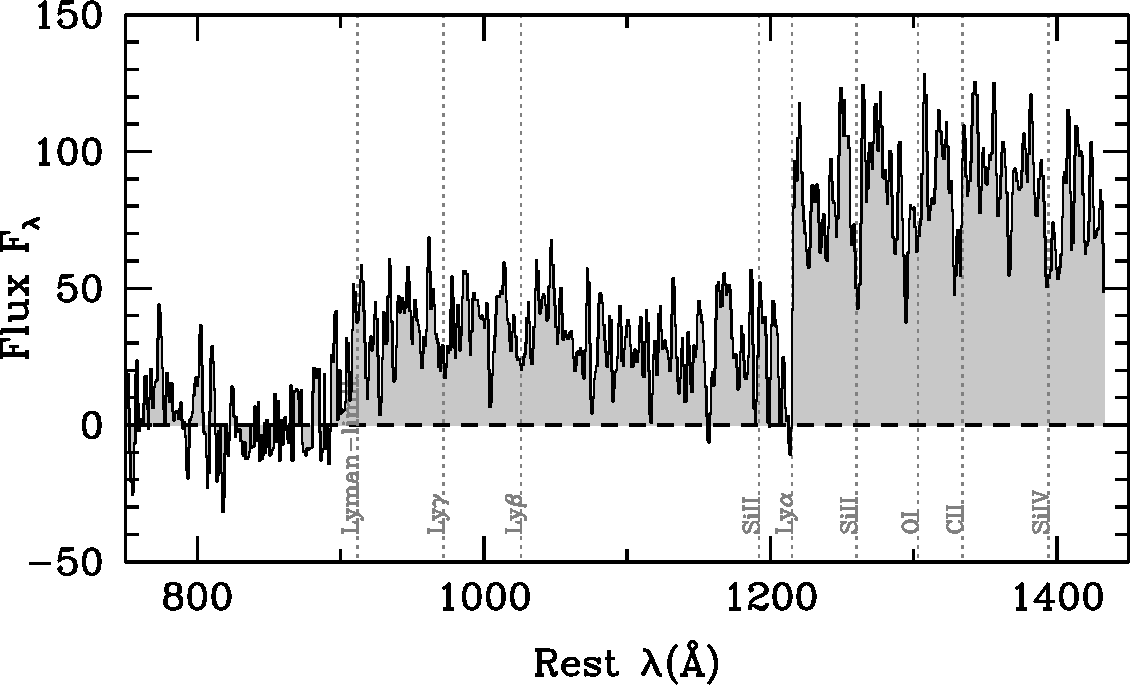}}
  \caption{Median stack of 11 spectra at $5.0 \leq z \leq6.6$ (median $z\sim5.6$) with Ly$\alpha$ in absorption.}
  \label{spec_abs}
\end{figure}

We present median stacked spectra of galaxies in our sample, normalized on the continuum at 1400\AA~rest-frame in Fig. \ref{spec_em} for galaxies with Ly$\alpha$ in emission and Fig. \ref{spec_abs} for Ly$\alpha$ in absorption. The continuum is detected in both cases, without significant emission below the Lyman limit at 912\AA, an indication that our sample selection and subsequent screening for low redshift interlopers is efficient in keeping only objects at $5.0\leq z \leq 6.6$. The stack of emission line galaxies corresponds to galaxies with a median $M_{FUV}$=-20.50, fainter than for the absorption stack with a median $M_{FUV}$=-20.78 (for galaxies with unknown FUV-magnitude an upper limit of -19.0 was used). The most prominent line in both cases is  Ly$\alpha$, with EW$_0(Ly\alpha)\simeq-100$\AA~in the emission spectrum (negative values of EW correspond to emission lines, positive to absorption), and EW$_0(Ly\alpha)\simeq5$\AA~in the absorption spectrum. There are only weak traces of absorption lines in the stack of Ly$\alpha$ emitting galaxies, while on the stack of spectra with Ly$\alpha$ absorption the brighter luminosities allow to identify  in absorption the Lyman series Ly$\gamma-972\AA$, Ly$\beta-1026\AA$, Ly$\alpha-1215\AA$, as well as SiII$-1260\AA$, OI$-1303\AA$ CII$-1334\AA$ and SiIV$-1394$\AA. We defer the comparison of the spectral properties of star-forming galaxies at $z\sim5.6$ to the properties at lower redshifts to a future paper.

\subsection{Main sequence of star-forming galaxies at $z\sim5.6$}

\begin{figure}[h]
  \resizebox{\hsize}{!}{\includegraphics{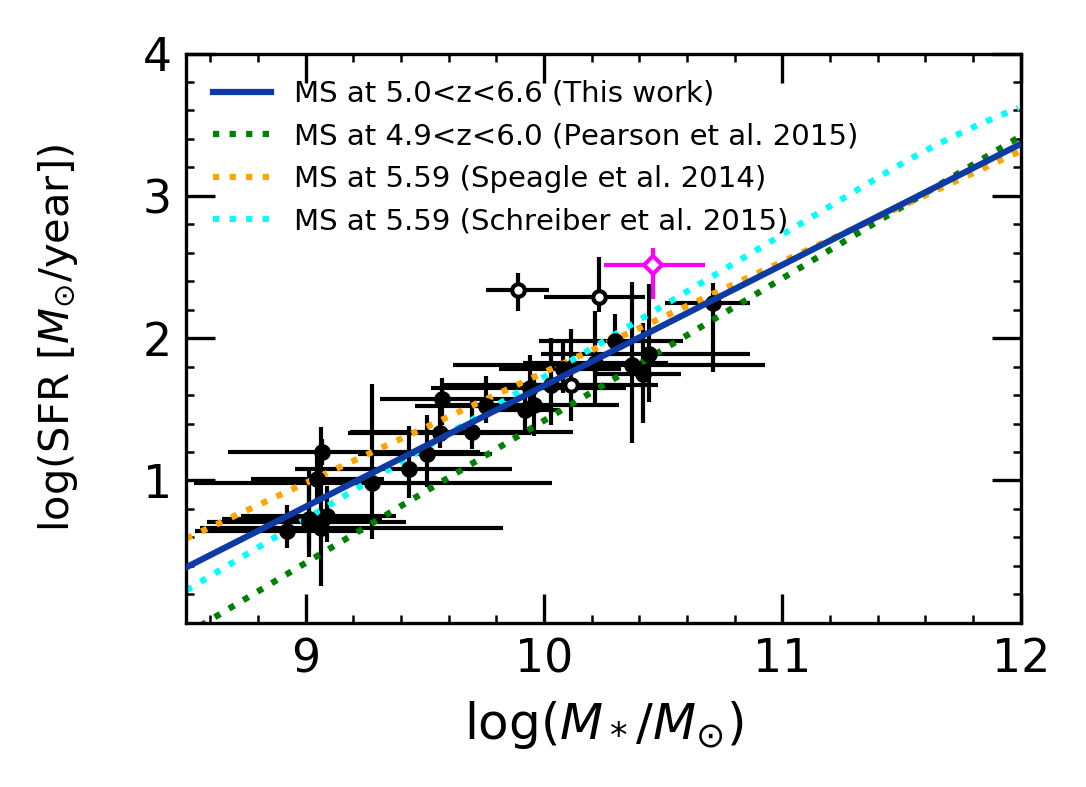}}
  \caption{SFR - $M_*$ diagram of our sample. The solid blue line is a fit to our data, representing the main sequence of star forming galaxies at these redshifts. The dotted cyan, orange and green lines are extrapolations of the main sequence at our median redshift from \cite{speagle_highly_2014, schreiber_herschel_2015, pearson_main_2018} . Filled circles are galaxies with reliable photometry and open circles are galaxies with possible contamination from bright nearby objects. The magenta diamond is the AGN identified in our sample.}
  \label{sfr_mass}
\end{figure}

The distribution of galaxies in the  SFR-stellar mass diagram is shown in Fig. \ref{sfr_mass}. Over the stellar mass range $9< \log{M_*/M_{\odot}}< 10.5$, SFR and stellar mass are tightly correlated, extending the existence of a 'main sequence' for star-forming galaxies to $5<z<6.6$ \citep[e.g.][]{elbaz_reversal_2007,whitaker_star-formation_2012,whitaker_constraining_2014,tasca_evolving_2015, santini_star_2017,pearson_main_2018}.  We fit the distribution with the a simple power law and find that $\log{}SFR(M_{\odot}yr^{-1}) \propto \alpha \times \log{M_*/M_{\odot}}$ with $\alpha= 0.85 \pm 0.05$ at $z\sim5.5$.

The relation is quite tight with a dispersion 0.13 dex around the mean, and all galaxies in our sample lie close to this main sequence, except for a few galaxies with  photometry affected by the contamination of nearby objects.

\begin{figure}[h]
  \resizebox{\hsize}{!}{\includegraphics{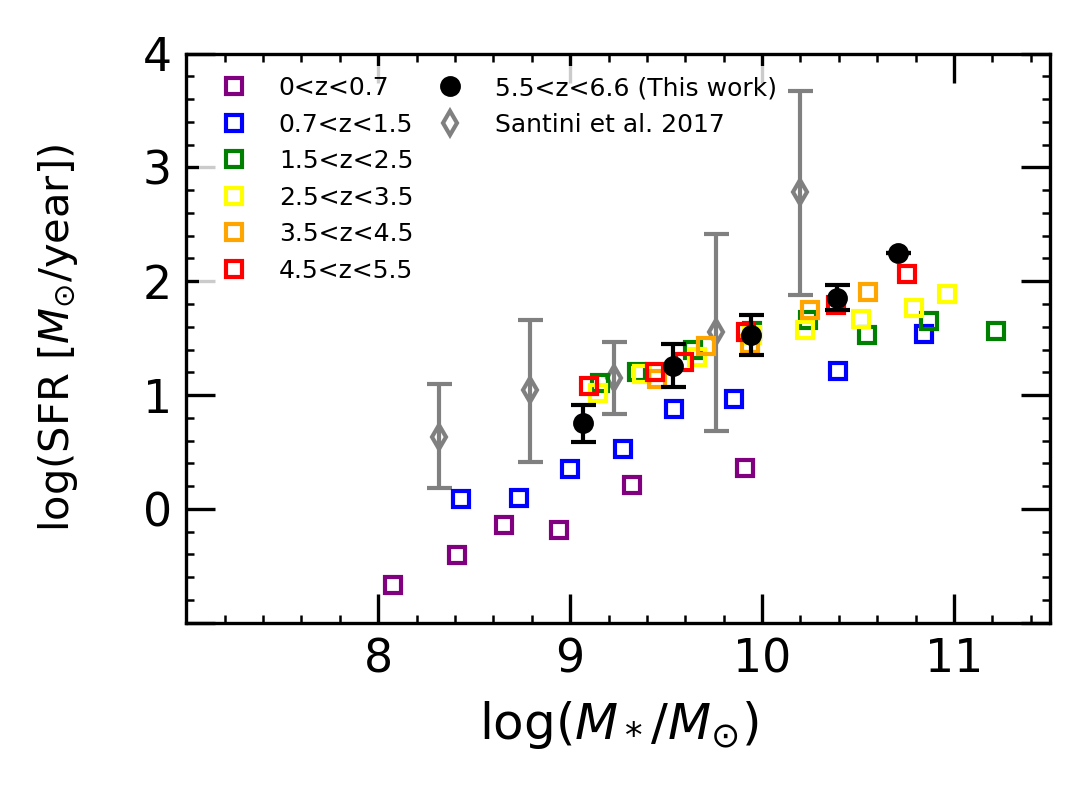}}
  \caption{Main sequence of VUDS galaxies at different redshifts. The colored squares show the median SFR in mass bins from \cite{tasca_evolving_2015}. The black circles are median SFRs from this work at 5.5<z<6.6.}
  \label{ms_plot}
\end{figure}

Previous studies show that at higher masses the main sequence has a turn-over observed at z$\sim$2, which however becomes less prominent at higher redshifts \citep[e.g.][]{whitaker_constraining_2014,tasca_evolving_2015,santini_star_2017,pearson_main_2018}. We show in Fig. \ref{ms_plot} the median SFR of galaxies in different mass bins for redshift ranges from 0 to 6.6 from VUDS. The data for z<5.5 is taken from \cite{tasca_evolving_2015} and the data in the last redshift bin 5.5<z<6.6 is from this work. Since we use consistent methods to derive SFR and stellar mass, we are able to extend the previous results from VUDS to higher redshifts.

The turn-over in the MS relation is clearly observed at z<3.5 but it seems to disappear at higher redshifts. Essentially all our galaxies lay very close to the linear main sequence, suggesting that the majority of them are still star-forming and we do not observe a significant turn-over at high mass in the MS, as would be expected if star-formation in massive galaxies was starting to be quenched. However, we find that a few  individual galaxies are slightly below the MS at masses $\log{M_*/M_{\odot}}>10.3$. Hence, at the highest masses, quenching processes may  just be starting to be at work at these redshifts.

As already shown in \cite{tasca_evolving_2015}, the normalization in SFR of the main sequence rapidly evolves up to $z\sim2.5$. At higher redshifts the normalization does not seem to evolve significantly (as shown on Fig. \ref{ms_plot}) and our data confirm that up to z$\sim6.6$ the normalization stays roughly constant.

\subsection{Specific star formation rate}

\begin{figure*}[h]
\centering
  \includegraphics[width=11cm]{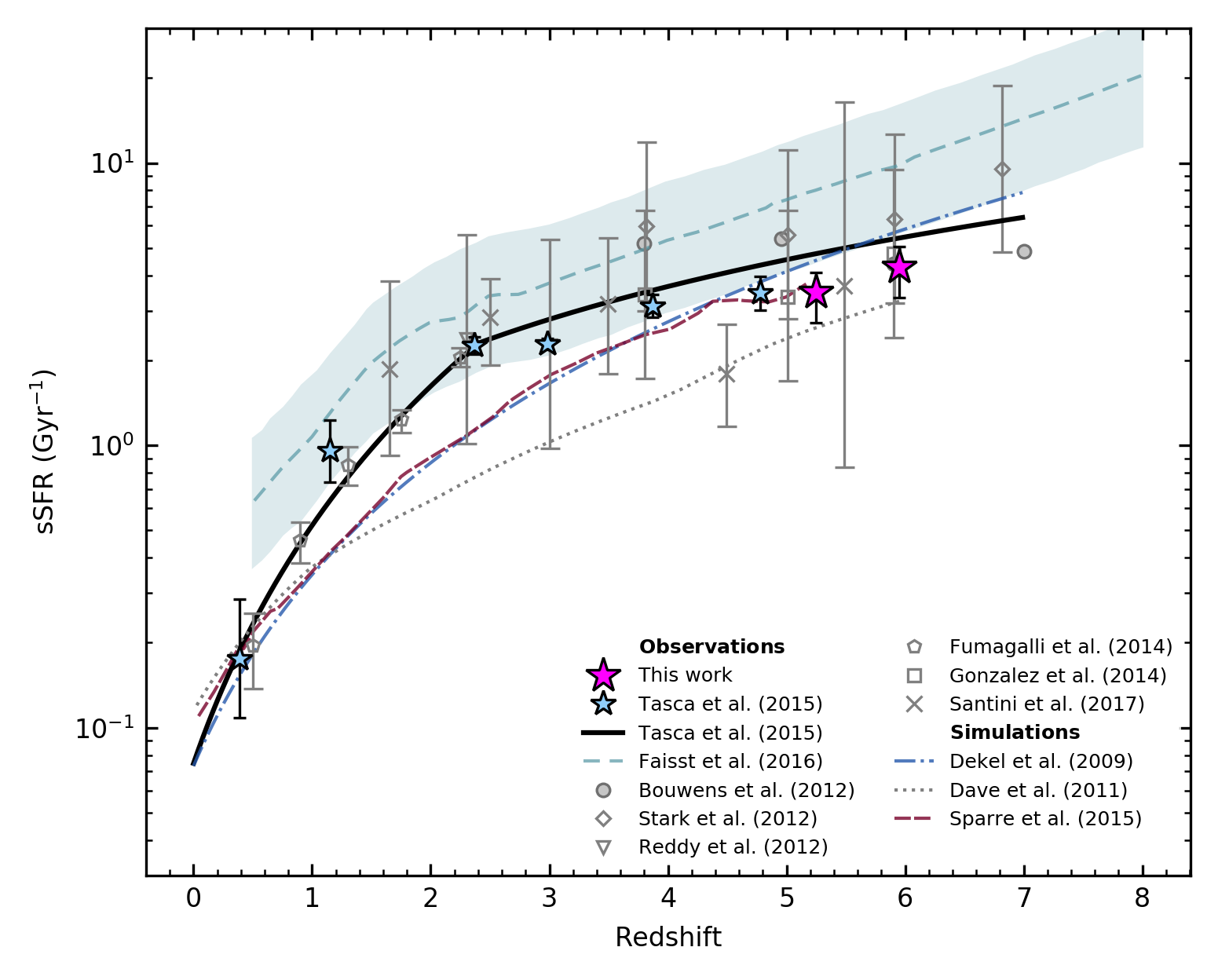}
  \caption{Redshift evolution of the sSFR. Results of various observations and numerical simulations are shown. The magenta star shows the result of this work, an extension of \cite{tasca_evolving_2015} work (light blue stars). The solid line shows the fit to the previous results from VUDS \citep{tasca_evolving_2015}, the shaded area and the dashed line shows the results, coming from the observations of EW(H$\alpha$) in COSMOS.}
  \label{ssfr_z}
\end{figure*}

We also extend the results of \cite{tasca_evolving_2015} on sSFR obtained with VUDS to the $5.0<z<6.6$ redshift range. It was previously shown that the sSFR evolution flattens at redshifts higher than z$\sim$2.4  \citep{tasca_evolving_2015,Faisst2016}. We compute the median sSFR of our sample using a lower stellar mass limit of $M_*>10^{10}M_{\odot}$. We find that $\log{}sSFR(Gyr^{-1})=-8.37\pm0.08$ and $-8.46\pm0.09$ at $5.0<z<5.5$ and $5.5<z<6.6$ respectively. Our highest redshift bin is only slightly higher than $\log{}sSFR(Gyr^{-1})=-8.46\pm0.06$ found by \cite{tasca_evolving_2015} at $4.5<z<5.5$. We conclude that the flattening of the sSFR continues in the redshift range up to $z~6.6$.

Our results  are shown in Fig. \ref{ssfr_z}. Over the redshift range of our sample, they are in excellent agreement with the latest results from HST Frontier Fields \citep{santini_star_2017}, with semi-analytical models based on gas accretion via cold streams \citep{dekel_cold_2009} and with the Gadget-2 and Illustris numerical simulations \citep{dave_galaxy_2011, sparre_star_2015}. 

These simulations, however, do not reproduce the observed sSFR evolution going from fast and steep at $z<2$ to slow and almost flat at $z>2$. Our results together with previous results from VUDS and other surveys therefore severely challenge our understanding of the processes driving the evolution of the sSFR through cosmic time.

\subsection{UV-continuum slopes}
\label{betaslopes}

\begin{figure}[h]
  \resizebox{\hsize}{!}{\includegraphics{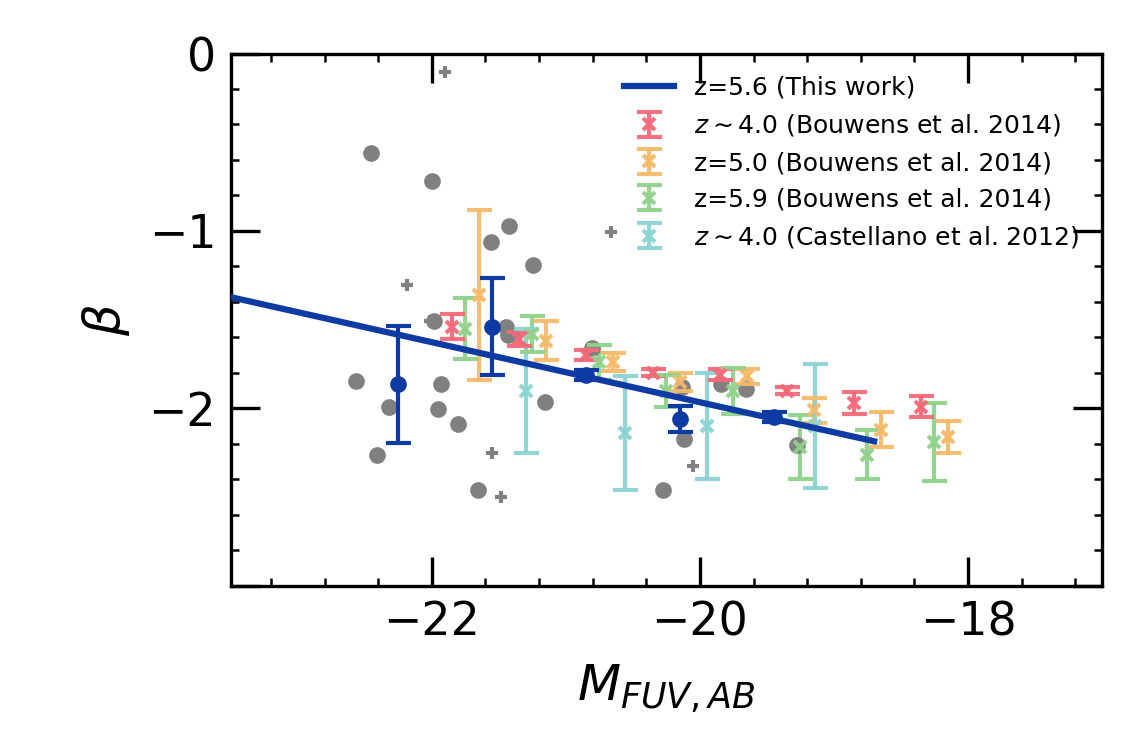}}
  \caption{$\beta$-slope vs. FUV-magnitudes of our sample. Grey circles indicate the individual measurements for the galaxies with most reliable photometry and grey crosses for the remaining ones. Blue circles are biweight means in bins of 0.7 mag size. The straight blue line is the linear fit to the biweight mean values. The colored crosses are values from \cite{bouwens_uv-continuum_2014} and \cite{Castellano2012} at different redshifts.}
  \label{beta_FUV}
\end{figure}

\begin{figure}[h]
  \resizebox{\hsize}{!}{\includegraphics{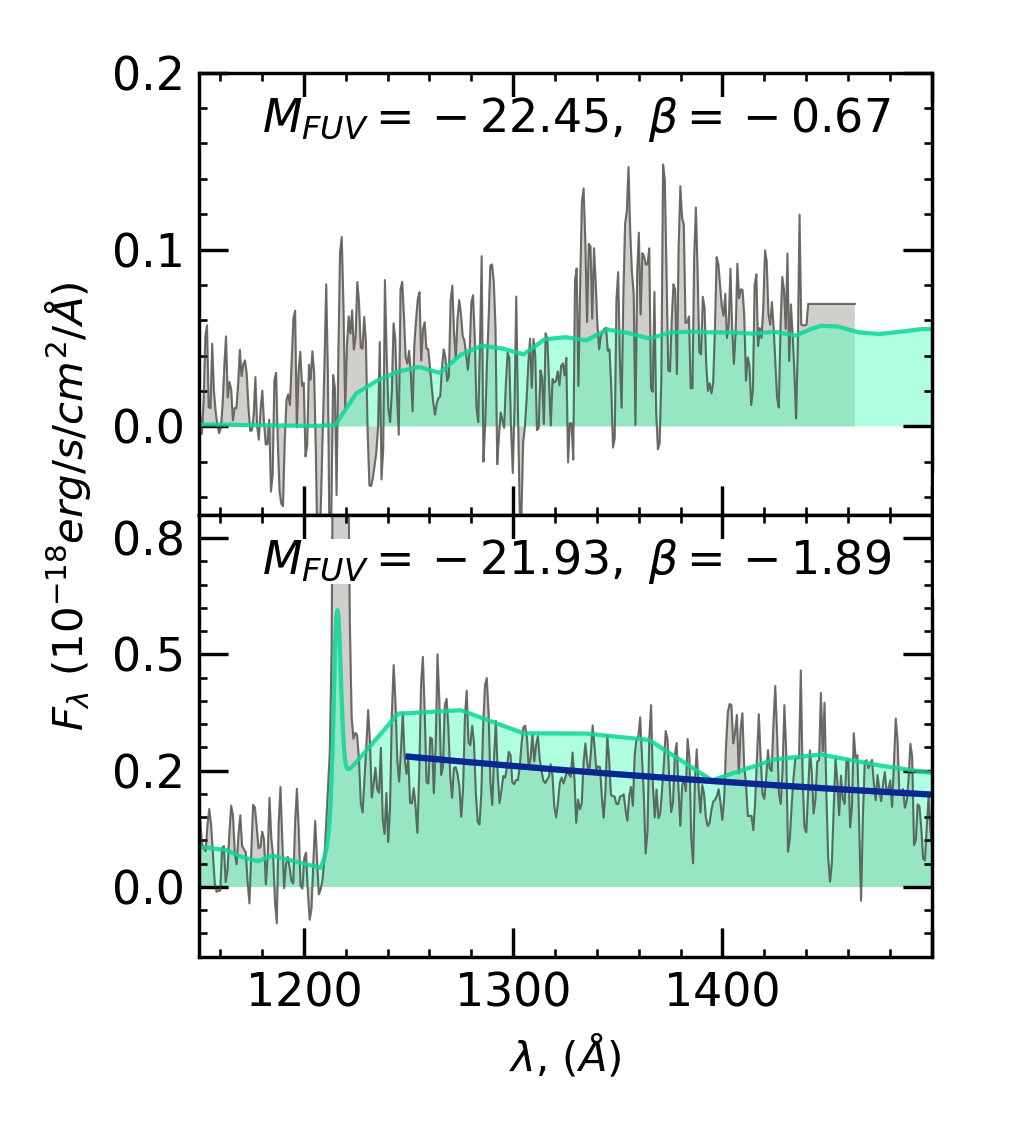}}
  \caption{Spectra of UV bright galaxies with flat and steep UV-continuum slopes (id=528295041, 520180097 at z=5.487 and 5.1378, respectively). The rest frame spectra are plotted in grey and the best SED template in light green. The solid blue line shows our fit to the continuum on spectra.}
  \label{spectra_beta_slopes}
\end{figure}

Using SED fitting we derive the UV-continuum slopes $\beta$. We fit the region from 1490 \AA~to 2350 \AA~on the best fit template with a power law $f_{\lambda}\sim\lambda^{\beta}$ \citep[e.g.][]{meurer_dust_1999}. For galaxies with high SNR for UV continuum, we fit templates to spectra as well as photometry and find consistent results (see Fig. \ref{spectra_beta_slopes}). In Fig. \ref{beta_FUV} we show the relation between the UV-continuum slope $\beta$ and the rest frame FUV absolute magnitudes for galaxies with the most reliable photometry.

We see a tentative decrease in the biweight mean $\beta$ with $M_{FUV}$, similar to the results of \cite{bouwens_uv-continuum_2014}. We also note, that most of the  $\beta$ measurements lay below the average values of \cite{bouwens_uv-continuum_2014} at z=4 and z=5, which indicates, that the galaxies in our sample are on average less dusty than galaxies LBGs at z<5, although the results of \cite{Castellano2012} at $z\sim4$ have steeper slopes and are in a better agreement with our results. For $M_{FUV}>-22$ we observe a steepening of the continuum slopes to fainter FUV magnitudes with a similar slope as in \cite{bouwens_uv-continuum_2014}.

However, at brightest magnitudes we observe a large scatter and a possible change of this behavior with some galaxies having very steep $\beta$ slopes. However, we also observe a galaxy with the reddest color and flat $\beta$ slope in the same bin. This galaxy is shown in Fig. \ref{spectra_beta_slopes} and has a robust estimate of $\beta=-0.56\pm0.05$ from the photometric fit and $\beta=-0.67\pm0.23$ from the fit to spectroscopy. This galaxy, therefore, is not an outlier. We conclude, that the scatter of $\beta$ is large and the brightest galaxies in this redshift range seem to be diverse in their spectral slopes, which may indicate different dust properties. The properties of galaxies are correlated with their age: the oldest galaxies  had enough time to build up dust and appear redder in their continuum, while younger, recently formed galaxies lack dust and appear bluer. 

\subsection{Age and formation redshift distribution}

Another property, which can be inferred from the SED fitting, is the age of the dominant stellar population(s). The median age of galaxies in the sample is 0.35 Gyr and the redshift of formation varies from 5.5 for the youngest galaxies at z$\sim5$ to 10.7 for the oldest galaxies at the highest redshifts. Over 90\% of the galaxies in our sample have a redshift of formation z>6.0, and therefore might have contributed to the reionization of the Universe since they were born, if they had non-negligible Lyman continuum escape fractions.

\section{Luminosity functions}
\label{LFs}

\subsection{1/V$_{max}$ method}
\label{vmax_method}

We use the  1/V$_{max}$ method \citep{schmidt_space_1968} to determine luminosity functions of our sample.  Each galaxy is weighted as

\begin{equation}
\label{weights}
w_{i}=\frac{1}{TSR*SSR},
\end{equation}

where TSR is the target sampling rate and SSR is the spectroscopic success rate.

\begin{figure}[h]
  \resizebox{\hsize}{!}{\includegraphics{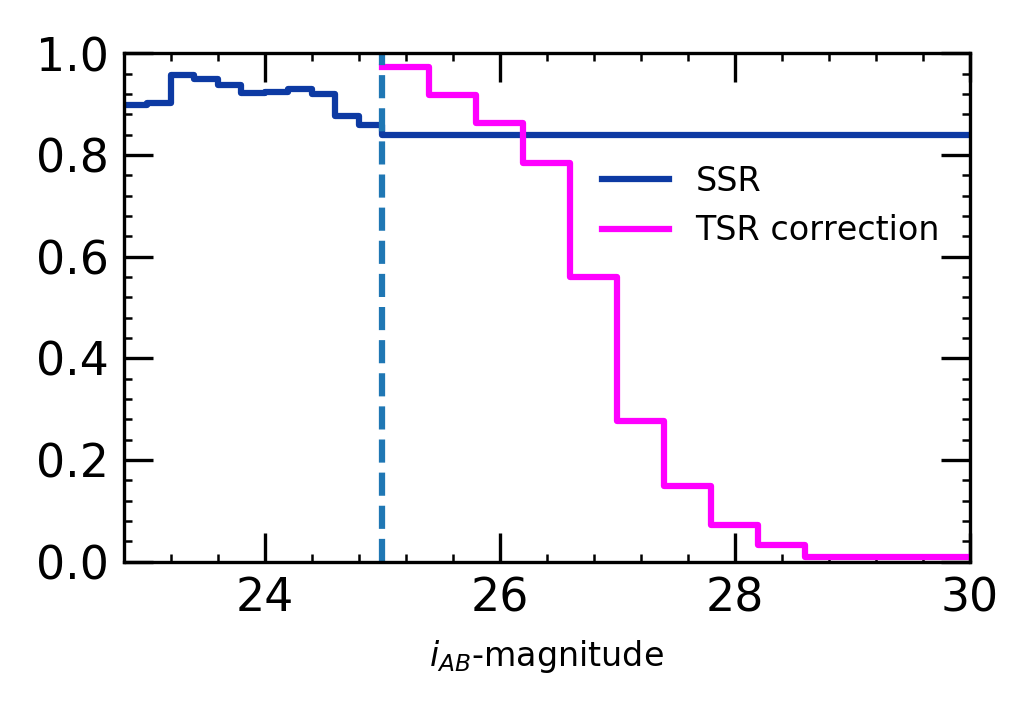}}
  \caption{The SSR as a function of $i$ magnitude and the corrections applied to TSR at magnitudes $i>25$, below the completeness limits.}
  \label{SSR_plot}
\end{figure}

The galaxies included in our sample are drawn from different selection criteria, therefore the TSR will depend on the selection criteria used. The TSR is also different for galaxies with $z>4$, because galaxies at these redshifts were prioritized targets. The parent population of galaxies is known for magnitudes $i<25$, a magnitude where the parent catalog is complete. At faint magnitudes we have to apply additional correction to the TSR based on the ratio of $N_{cat}/N_{exp}$, where $N_{cat}$ is the number of galaxies in the parent catalog and $N_{exp}$ is the expected number of galaxies at fainter magnitudes, if the catalogue were complete. We derive this number by extrapolation from the $i$ magnitude distribution in the photometric catalog. These corrections are shown in \ref{SSR_plot}.

The underlying parent population of serendipitous galaxies is not known. We assume that all faint galaxies which fall on the slit area are observed, and we estimate the TSR as the ratio of the area covered by slits to the whole observed area, which is equal to $\sim0.2\%$. A few galaxies with $i>25$ from the parent catalog are also observed, if they fall into the slits. We treat them as serendipitous and use the same TSR.

To summarize, we evaluate the TSR for galaxies with $z>4$ for the bright and faint subsamples corresponding to each selection criteria and we multiply the TSR of faint galaxies by additional corrections (see Fig. \ref{SSR_plot}). All TSR used in this paper are shown in Table \ref{TSR_table}.

The SSR should not depend on the selection criteria, but it depends on the  $i$-band magnitude. We evaluate it for the whole sample of galaxies with $i<25$ as a function of magnitude. The SSR of fainter galaxies is more uncertain, as it starts to depend on the strength of the Ly$\alpha$ line, rather than on the brightness of the continuum. Indeed, the strong LAE have fainter FUV-continuum, but higher probability to be detected. We therefore ignore the dependence on $i$-band magnitude for the fainter sub-sample and assume a constant SSR based on  Ly$\alpha$  emission  detection limits. The resulting SSR is shown in Fig. \ref{SSR_plot}.

\begin{table}
\caption{The target sampling rates of different selection criteria}
\label{TSR_table}                         
\begin{tabular}{c c c} 
\hline      
\hline             
Criterion & $i$ & TSR (\%)\\   \hline 
$z_{phot}>4.0$ & $i<25$ & 3.6 \\
Color-color criteria & $i<25$ & 2.8 \\
 & $i>25$ & 3.4 \\
Narrow band selection & $i<25$ & 2.5 \\
 & $i>25$ & 3.6 \\
Serendipitous & $i>25$ & 0.2 \\ 
\hline 
\end{tabular}
\end{table}

After assigning the weights, as described above, we determine the luminosity function as

\begin{equation}
\label{Vmax}
\phi(M)=\frac{1}{dM}\sum_{i=1}^{N_{gal}}\frac{w_i}{V_{max,i}},
\end{equation}

Where $M$ is the FUV magnitude or the $Ly\alpha$ luminosity in log, $\phi(M)$ is the number density in magnitude or luminosity bin, $dM$ is the bin size, $N_{gal}$ is the total number of galaxies and $V_{max,i}$ is the maximum comoving volume where the i-th galaxy can be observed. For the bright subsample of galaxies we determine the volume $V_{max,i}$ by using the limits on $i$ magnitude. For the faint galaxies ($i>25$), the volume in which they are observed depends mainly on the flux of Ly$\alpha$ line. 

We calculate Poisson errors of our results as well as errors induced by the weights. For the latter, we calculate the luminosity function using the upper and lower limits of the weights, which are defined by the estimated errors on the weights.

\subsection{UV luminosity function}
\label{UVLF_section}

Before determining the UV luminosity function we investigate how uncertainties of the observed magnitudes propagate into the uncertainty of FUV magnitudes determined with LePhare. We take a set of observed magnitudes of each galaxy and then sample 500 new magnitude sets, assuming Gaussian errors on the measured flux. We use these magnitude sets to recompute the absolute magnitude using the same method and compare the new values with the $M^0_{FUV}$ -- the best estimate of the absolute magnitude of a galaxy. We obtain, in this way, a distribution of $\Delta{M_{FUV}}$ for each individual galaxy.

The inspection of these distributions shows that galaxies with the smallest number of photometric detections (2-3) have the largest uncertainties on $M_{FUV}$. These galaxies are only detected in the bands where the emission lines are located, such as the i-band or z-band for Ly$\alpha$ and IRAC bands for [OIII] and H$\alpha$ lines. Therefore, for these galaxies, the estimation of  $M_{FUV}$ strongly depends on the assumptions made about the strength of the emission lines. We introduce these galaxies into the luminosity function by weighting them with the probability for each of them to be inside each absolute magnitude bin. To compute this probability, we normalize the distribution of $\Delta{M_{FUV}}$, obtaining the probability distribution of the absolute magnitude and integrate this distribution between the bin limits. 

Although the distribution of $\Delta{M_{FUV}}$ varies slightly from galaxy to galaxy, the average uncertainty remains almost the same for a given photometric set used, which is different depending on the field (as discussed in Sect. \ref{data}). The average uncertainties are 0.07, 0.04, 0.05 for COSMOS, VVDS02h and ECDFS fields respectively. For a few galaxies in ECDFS field with photometry from TENIS, the average uncertainties are larger ($\sim0.14$), due to a small number of bands. 

\begin{figure}[h]
  \resizebox{\hsize}{!}{\includegraphics{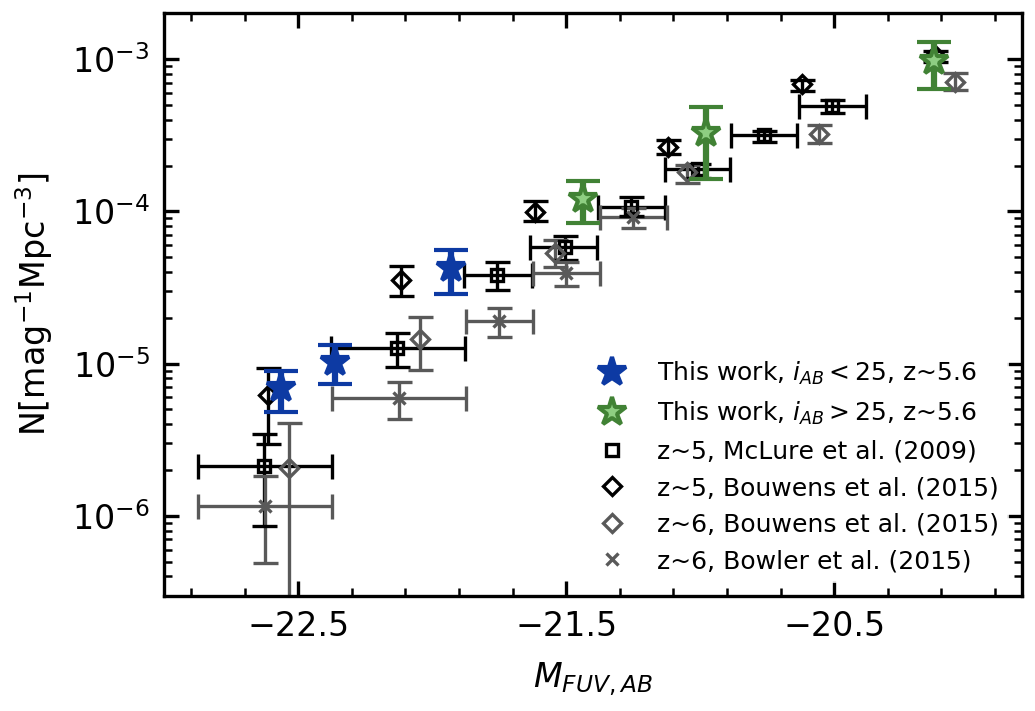}}
  \caption{The  UV luminosity function at $5.0<z<6.6$, at the median redshift $z=5.6$. The blue stars are UV luminosity function estimations drawn from the bright $i<25$ sample, the green stars are from the faint sample with completeness correction as described in \ref{vmax_method}. The black open symbols are UV luminosity function from the literature at z$\sim$5 and the grey ones at z$\sim$6.}
  \label{UVLF}
\end{figure}

After examining the quality of $M_{FUV}$ magnitudes we proceed to determine the UV luminosity function of our sample. We present our results in Fig. \ref{UVLF} and Table \ref{table_LF_data}. We compare our results with luminosity functions reported in the literature at z=5 and z=6 \citep{mclure_luminosity_2009,bouwens_uv_2015,bowler_galaxy_2015} and find a good agreement within error bars, our luminosity function being closest to the z=5 luminosity function of \cite{bouwens_uv_2015}.

\begin{figure}[h]
  \resizebox{\hsize}{!}{\includegraphics{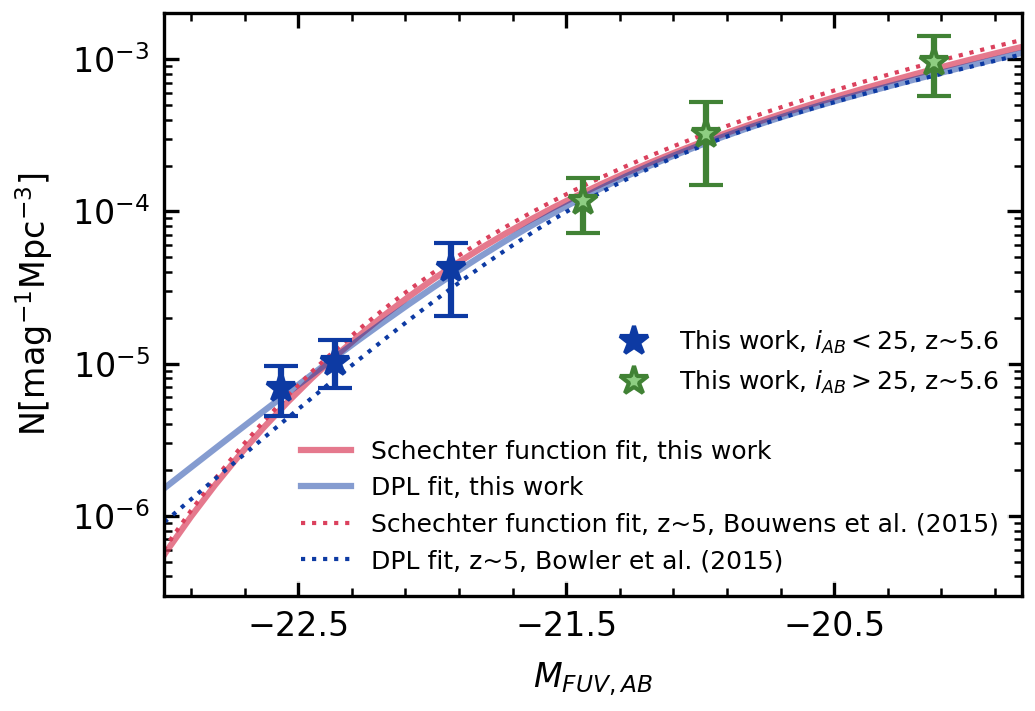}}
  \caption{Fit of the UV luminosity function with a DPL (blue solid line) at $z=5.6$ compared to a Schechter function (red solid line). The dotted lines of respective colors are fits from the literature. The filled stars are the same as in Fig. \ref{UVLF}}
  \label{UVLFfit}
\end{figure}

\cite{bowler_galaxy_2015} reported that the bright end of the luminosity function at $z\sim6$ has a higher number density than expected from a classical luminosity function \cite{Schechter1976} shape, and is better represented by a double power law (DPL). We try to fit two functional forms of the luminosity function -- a standard Schechter function form and a DPL. We fit the parameters of the luminosity function in these two representations with a MCMC method implemented within the python package {\texttt pymc}. Because our sample is mostly built from bright star forming galaxies, our measurements of the faint end are not well constrained, while we set strong constraints on the bright end. In order to fit the luminosity function, we set the faint end slope to values from the literature ( $\alpha=-1.76$ from \cite{bouwens_uv_2015} and $\alpha=-2.0$ from \cite{bowler_galaxy_2015}). We also set $\phi^*$, when fitting with a DPL, due to our small sample. Our results are shown in Fig. \ref{UVLFfit} and listed in Table \ref{table_LF_fit}.

Both a Schechter or a DPL fit represent well our data at all magnitudes. However, the reduced $\chi^2$ of the fit with DPL is lower (see Table \ref{table_LF_fit}) and for the bright sample ($i<25$), the reduced $\chi^2$ of DPL is even 2.5 times lower compared to the Schechter function fit. Since the parent catalogue is complete for the bright sub-sample, we expect these data to be the most reliable. We therefore conclude that the luminosity function is at z$\sim5.6$ can be better represented by a DPL.
 
\begin{table}
\caption{UV and $Ly\alpha$ luminosity function measurements.}
\label{table_LF_data}    
                 
\begin{tabular}{c c c c}       
\hline 
\hline             
$\log_{10}{L_{Ly\alpha}}(erg/s)$ &  $\phi(10^{-4}Mpc^{-3})$  & bin size & $N_{gal}$ \\   \hline 
 42.06 & $44.6_{- 8.6}^{+ 10.1}$ & 0.35 &  10 \\
 42.48 & $27.4_{- 5.5}^{+ 4.3}$ & 0.35 &  15 \\
 42.69 & $20.4_{- 4.9}^{+ 5.9}$ & 0.35 &  6 \\
 43.12 & $6.18_{- 4.0}^{+ 2.0}$ & 0.35 &  3 \\
 43.59 & $0.06_{- 0.02}^{+ 0.24}$ & 0.35 &  1 \\
\hline 
$M_{FUV}$ &  $\phi(10^{-4}Mpc^{-3})$  & bin size & $N_{gal}$ \\   \hline 
-22.56 & $0.07_{- 0.02}^{+ 0.03}$ & 0.3 & 1 \\
-22.36 & $0.10_{- 0.03}^{+ 0.04}$ & 0.4 & 2 \\
-21.93 & $0.43_{- 0.22}^{+ 0.20}$ & 0.5 & 3 \\
\hline
-21.44 & $1.16_{- 0.44}^{+ 0.48}$ & 0.6 & 6 \\ 
-20.98 & $3.2_{- 1.7}^{+ 2.1}$ & 0.6 & 4 \\
-20.12 & $9.6_{- 3.8}^{+ 4.6}$ & 0.6 & 4 \\
\hline
\end{tabular}
\end{table}

\begin{table*}
\caption{Parametric fitting of the UV and $Ly\alpha$ luminosity functions. }
\label{table_LF_fit}    
\centering                       
\begin{tabular}{c c c c c c c c c}       
\hline \hline
LF &  $\alpha$ & $M^*$  & $\phi^*/10^{-4}$ & $\beta$ & $\log{}SFRD_{uncorr}$ & $\log{}SFRD_{corr}$ &  $\chi^2_{whole}$ & $\chi^2_{bright}$ \\ 
 &  & (mag) & (mag$^{-1}$Mpc$^{-3}$) &  & \multicolumn{2}{c}{(M$_{\odot}$yr$^{-1}$Mpc$^{-3}$)}  &  & \\ \hline 
UV\tablefootmark{a} & $-2.00$ & $-21.43_{- 0.10}^{+ 0.13}$ & $2.5$ & $-4.52_{-0.48  }^{+ 0.49 }$ & $-1.63_{-0.08  }^{+ 0.06 }$ & $-1.34_{-0.08  }^{+ 0.06 }$ & 0.72 & 0.37 \\
UV\tablefootmark{b} & $-1.76$ & $-21.10_{- 0.15}^{+ 0.13}$ & $7.1_{-2.5 }^{+ 3.2}$ & - & $-1.63_{-0.16  }^{+ 0.13 }$ & $-1.34_{-0.16  }^{+ 0.13 }$ & 1.18 & 0.86 \\ \hline
LF  & $\alpha$    & $\log_{10}{L^*_{Ly\alpha}}$  & $\log_{10}\Phi^*$ & $\log{}SFRD_{uncorr}$ &  $\log{}SFRD_{corr}$ & $\chi^2$ &    &\\ 
&  & $(erg s^{-1})$  & $(\Delta\log{L}^{-1}$Mpc$^{-3})$ &  \multicolumn{2}{c}{(M$_{\odot}$yr$^{-1}$Mpc$^{-3}$)} & & &\\ \hline 
Ly$\alpha$\tablefootmark{b} & $-1.69$ & $43.00_{- 0.12}^{+ 0.09}$  & $-3.21_{-0.10 }^{+ 0.12}$ & $-2.02_{-0.08  }^{+ 0.07 }$ & $-1.40_{-0.08  }^{+ 0.07 }$ & 4.95  &   &\\
Ly$\alpha$\tablefootmark{b} & $-1.76$ & $43.03_{- 0.12}^{+ 0.09}$  &  $-3.30_{-0.11 }^{+ 0.12}$& $-2.02_{-0.08  }^{+ 0.07 }$ & $-1.40_{-0.08  }^{+ 0.07 }$ & 5.78  &  &\\
Ly$\alpha$\tablefootmark{b} & $-2.00$ &  $43.15_{- 0.15}^{+ 0.12}$ & $-3.63_{-0.15 }^{+ 0.18}$ & $-2.05_{-0.09  }^{+ 0.08 }$ & $-1.43_{-0.08  }^{+ 0.07 }$ & 21.9  &  & \\
\hline
\end{tabular}
\tablefoot{
\tablefoottext{a}{parameterized as DPL}
\tablefoottext{b}{parameterized as Schechter function}
}
\end{table*}

\subsection{Ly$\alpha$ luminosity function}

\begin{figure}[h]
  \resizebox{\hsize}{!}{\includegraphics{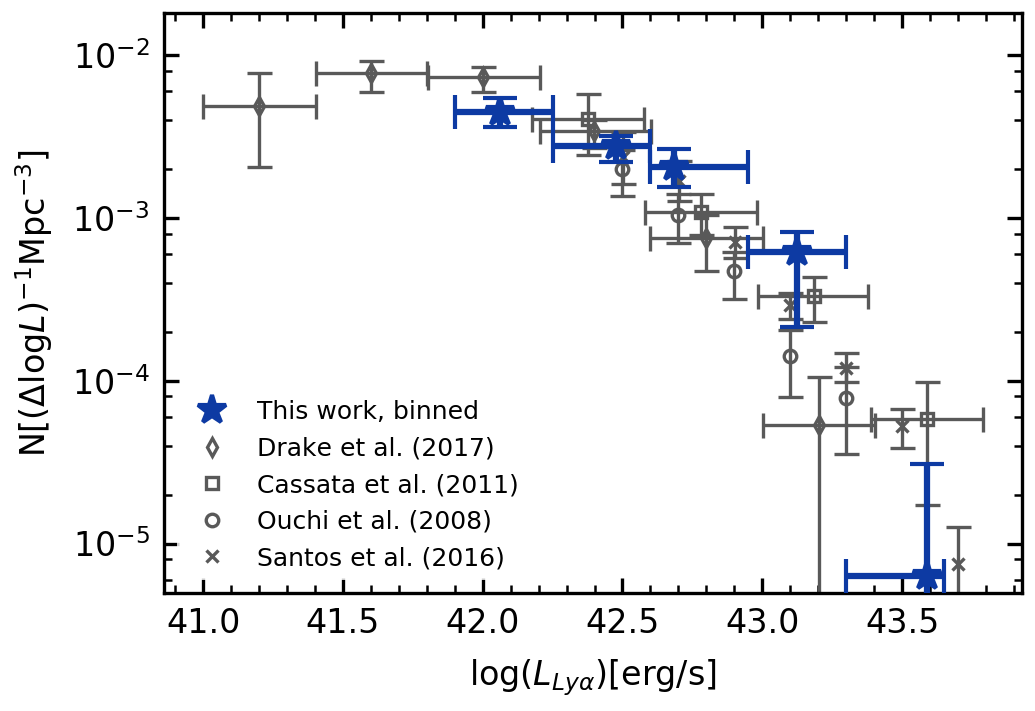}}
  \caption{Ly$\alpha$ luminosity function at the median redshift $z=5.6$ (blue stars). The open symbols are previous results from the literature.}
  \label{LyaLF_binned}
\end{figure}

\begin{figure}[h]
  \resizebox{\hsize}{!}{\includegraphics{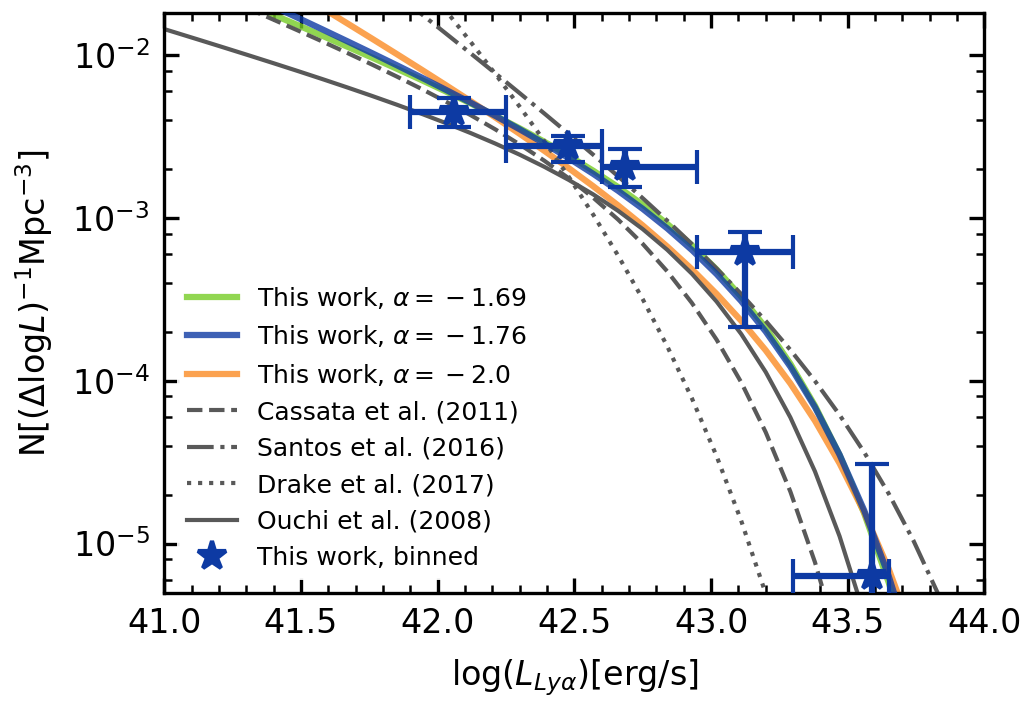}}
  \caption{Ly$\alpha$ luminosity functions fitted with a Schechter function at redshifts $5.0<z<6.6$. The colored solid lines are fits to our data with different faint end slopes, the grey lines are results from the literature. The filled stars are the same as in Fig. \ref{LyaLF_binned}.}
  \label{SchechterLyaLF}
\end{figure}

We measure Ly$\alpha$ fluxes manually, using the {\texttt splot} tool in {\sc IRAF}. We proceed in the following way: first, we interpolate the continuum flux at Ly$\alpha$ from the continuum levels redward of Ly$\alpha$ and measure the flux in the line above this level. Then, we place the continuum level 1$\sigma$ (RMS of continuum measurements) above and below the average value of the continuum redward from Ly$\alpha$, to estimate the errors of our measurements. We also measure the ratio of continuum flux red and blueward from Ly$\alpha$ for the galaxies without the emission line, but with a visible break in the continuum.

All fluxes are corrected for  slit losses. Slit losses in VVDS, a survey with a nearly identical observational setup to VUDS,  were extensively studied by \cite{cassata_vimos_2011} and we apply the same corrections. For the targeted galaxies, centered on the slits, the recovered flux is $\sim85\%$ and for the serendipitous objects the median value is $\sim55\%$.

We compute the  Ly$\alpha$ luminosity function as described in Sect. \ref{vmax_method} and present our results in Fig. \ref{LyaLF_binned} and Table \ref{table_LF_data}. Given the detection limits for the Ly$\alpha$ flux in spectra, we expect our sample to be complete up to $\log_{10}L_{Ly\alpha}(erg s^{-1})\sim42.0$.

The observed bright end of the luminosity function is in good agreement with \cite{cassata_vimos_2011} (for $4.55<z<6.6$) and \cite{santos_lyman-alpha_2016} (for LAE at z=5.7). On the bright end the number density decreases, but not as fast as reported from the MUSE deep fields  \citep{drake_muse_2017} or \cite{ouchi_subaru/xmm-newton_2008}. However, the uncertainty of the MUSE data is much higher at the bright end, because the small observed field is subject to strong cosmic variance, especially for the brightest galaxies \citep{moster2011}.

One of the important sources of uncertainty in the Ly$\alpha$ luminosity function is the faint end slope. Only recently some attempts to provide such constrains have been published, still very uncertain \citep{santos_lyman-alpha_2016, drake_muse_2017}.
Since our data are not constraining enough on the faint end slope we set it to  values from the literature as priors when fitting the Ly$\alpha$ luminosity function: $\alpha=-1.76, -2.00$, the same values, which we used for the UV luminosity function and $\alpha = -1.69$, as used in \cite{cassata_vimos_2011}. We also test a wide range of faint end slopes from  $\alpha=-1.5$ to -2.3. We use uniform priors on $L^{*}_{Ly\alpha}(erg s^{-1})$ and $\phi^*$ ($35<\log_{10}L^{*}_{Ly\alpha}<50$ and $-15<\log_{10}\phi^*<-1$) and run MCMC minimization to find the best fit of the Ly$\alpha$ luminosity function.

Results are given in Table \ref{table_LF_fit} and Fig. \ref{SchechterLyaLF}. As expected, uncertainties on the faint end slope lead to uncertainties on the Schechter function parameters $\phi^*$ and $M^*$ left free in the fit. As the slope $\alpha$ is set to steeper  values, one gets  a brighter  $L^{*}_{Ly\alpha}$ and a lower  $\phi^*$. 

The steep values of the faint end slope ($\alpha<-2.0$) do not agree well with our data and we could not obtain a satisfactory fit with them. The latest works \citep{santos_lyman-alpha_2016, drake_muse_2017} suggest values of faint end slope below -2.0, but already with a slope $\alpha=-2.0$ it becomes challenging to fit both the bright and faint bins in our data. We therefore use the value $\alpha = -1.69$ in reporting our final results.

\section{Star formation rate density}
\label{sfrd}

Using our UV and Ly$\alpha$ luminosity functions we proceed to determine the SFRD within the redshift range of our sample.

To calculate the SFRD we integrate the luminosity functions to compute the luminosity density. For Ly$\alpha$ emitting galaxies we integrate from $0.04\times{L^*_{Ly\alpha}}$ (with $\log{}L^*_{Ly\alpha}=43.0$ from our best estimate) to $\log_{10}L_{Ly\alpha}=44$. We then transform it to SFRD as:

\begin{equation}
\label{sfrd_lya_eq}
SFRD_{Ly\alpha}[M_\odot yr^{-1} Mpc^{-3}] = L_{Ly\alpha}[erg s^{-1}]/ 1.1\times 10^{42}.
\end{equation}

We use the same conversion factor as \cite{cassata_vimos_2011}, based on the ratio between $L_{Ly\alpha}$ and $L_{H\alpha}$ of \cite{brocklehurst_calculations_1971} and the conversion factor between SFR and $L_{H\alpha}$ from \cite{kennicutt_global_1998}.

We integrate the UV luminosity function from $M_{FUV}=-17.0$ down to $M_{FUV}$ corresponding $100 \times L^*_{FUV}$ \citep{madau_cosmic_2014}. We use $\kappa_{FUV}=2.5 \times 10^{-10} [M_{\odot} yr^{-1} L_{\odot}^{-1}]$ from \cite{madau_cosmic_2014} to convert $L_{FUV}$ to $SFRD_{FUV}$. For both luminosity functions our lower integration limits correspond to the same lower SFR value, enabling consistent comparison between the SFRD traced by UV and Ly$\alpha$. We correct $SFRD_{FUV}$ for dust extinction using our measurements of $\beta$ slopes and IRX-$\beta$ relation from \cite{meurer_dust_1999}. We obtain $A_{FUV}=0.72$. 

\begin{figure*}[h]
\centering
  \includegraphics[width=11cm]{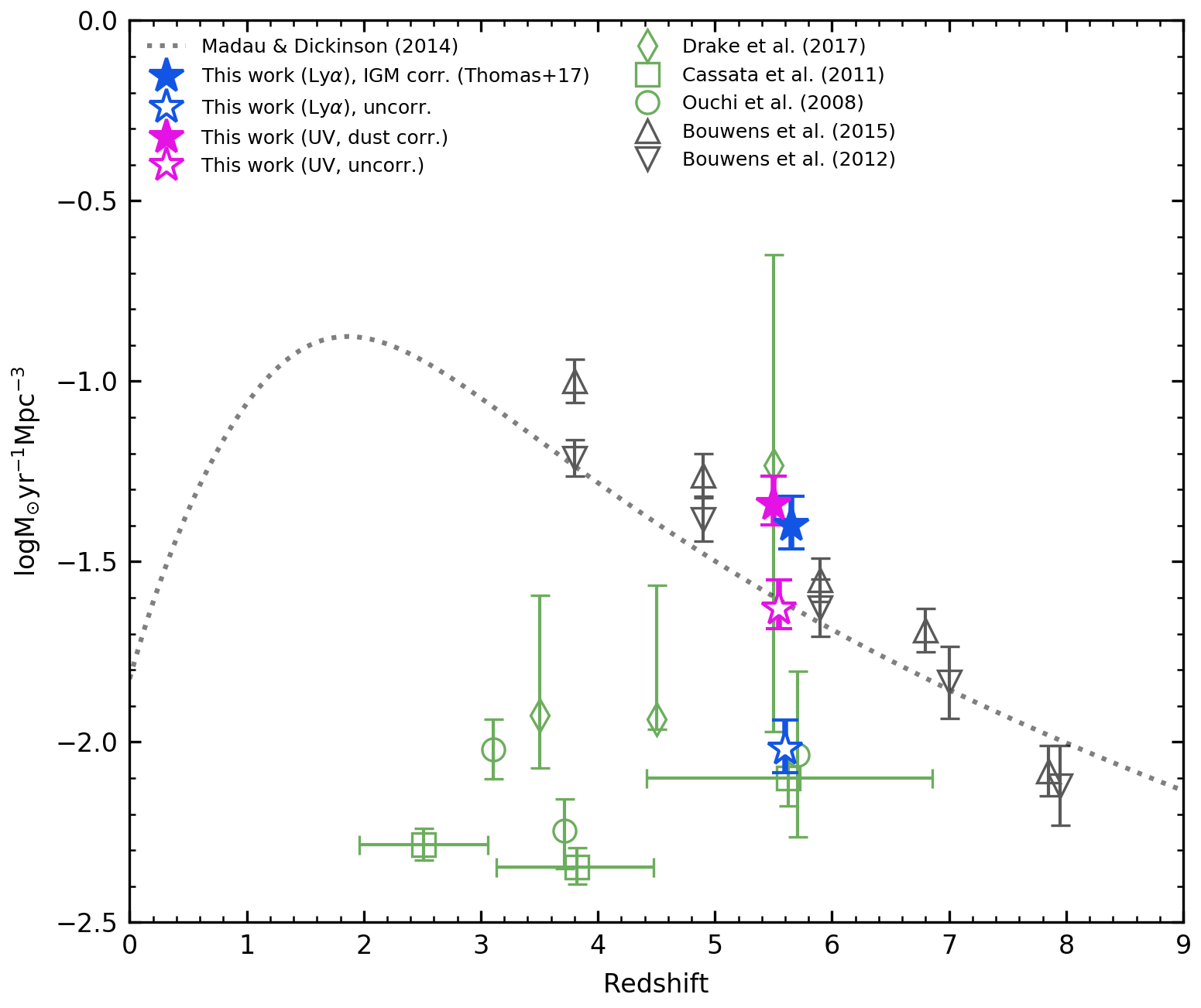}
  \caption{SFRD vs. redshift. The filled stars are results from this work. The light green points are Ly$\alpha$ luminosity function based measurements, the grey points are UV-based. The SFRDs from the literature are calculated from the luminosity functions using the same integration limits and conversion factors as in this work.}
  \label{SFRD}
\end{figure*}

Results are presented in Fig. \ref{SFRD} and Table \ref{table_LF_fit}. The error bars include the uncertainties on the fit and cosmic variance. Cosmic variance is calculated using the recipe of \cite{Driver2010} and is equal to 5\% given the geometry and population of our survey. 

We compute a UV derived SFRD of $\log{}SFRD_{UV}=-1.34_{-0.08  }^{+ 0.06 }$, obtained from the best fit to the luminosity function parametrized as a DPL. We obtain the same value from the fit with the Schechter function. The increase of number density of the bright end in case of DPL parameterization does not significantly change the estimate of the SFRD. Our result is slightly higher, by 0.2-0.3 dex, than the best fit to literature measurements as reported by  \cite{madau_cosmic_2014} (Fig. \ref{SFRD}), but is in agreement with \cite{bouwens_uv_2015} within error bars. 

Despite the uncertainty on the faint end slope, the SFRD$_{Ly\alpha}$ remains roughly constant within the error bars for the slopes in a range from $\alpha=-1.5$ to $\alpha=-1.85$, because when the faint end slope steepens, the normalization density decreases. For steeper values of $\alpha$ the normalization density starts to decrease faster and the best fit of the luminosity function falls below our measurements at the bright end. This leads to an underestimate of the contribution of  bright galaxies to the SFRD. Therefore we consider $\log{}SFRD_{Ly\alpha}=-2.02_{-0.08  }^{+ 0.07 }$,  obtained with $\alpha=-1.69$, to be our best estimate of the contribution from Ly$\alpha$ emitting galaxies to the SFRD.

This result is in agreement within error bars with previously published results for samples selected in completely different and independent ways \citep{ouchi_subaru/xmm-newton_2008, cassata_vimos_2011}. It differs by 0.76 dex from results obtained with MUSE observations of HUDF \citep{drake_muse_2017}, mainly due to the very steep faint end slope used by \cite{drake_muse_2017}. Our results, however, are in broad agreement when taking into account the large error bars of the \cite{drake_muse_2017} measurement.

The fact that we determine both the UV and Ly$\alpha$ luminosity functions  using the same sample of galaxies enables to get a robust constraint of the ratio $SFRD_{Ly\alpha}/SFRD_{UV}$. As discussed in \cite{Hayes2011}, this value can be an estimate of volumetric Ly$\alpha$ escape fraction $f^{Ly\alpha}_{esc}$. Using the same formalism we obtain robust estimate $f^{Ly\alpha}_{esc}=21\pm4\%$, same as \cite{Hayes2011} estimate $f^{Ly\alpha}_{esc}=21^{+19}_{-7}\%$ at z=5.6 (the value from a best fit of a compilation of measurements using previous works on UV and Ly$\alpha$ luminosity functions).

To obtain an estimate of the total number of Ly$\alpha$ photons emitted within a galaxy one has to correct Ly$\alpha$ flux absorption by the IGM. Observations of the Gunn-Peterson trough in high redshift quasars \citep{Fan2006} indicate, that more than half of the flux is absorbed by the IGM at our redshifts. The same results were obtained by \citep{thomas_vimos_2017} from VUDS at z<5.5. We estimate the IGM transmission of Ly$\alpha$ flux directly from the spectra in our sample using the same technique: we fit spectra with a range of SED templates combined with the prescription of \cite{Madau1995} on IGM transmission. Due to the degeneracy between IGM and dust attenuation, we limit the E(B-V) range to [0.0-0.1], therefore assuming a low dust content. We find that the mean Ly$\alpha$ transmission on our spectra is $Tr(Ly\alpha)=0.24$. If we correct the observed luminosity density $L_{Ly\alpha}$ by this value, we then obtain a corrected value of the Ly$\alpha$-derived SFRD $\log{}SFRD_{Ly\alpha}=-1.40_{-0.08  }^{+ 0.07 }$. This result is in excellent agreement with the UV-derived SFRD, within error bars (see Fig. \ref{SFRD}). It also indirectly indicates that our assumption on the low dust content of galaxies at these redshifts when computing the Ly$\alpha$-derived SFRD is broadly correct. We therefore show that using either the UV or Ly$\alpha$ luminosity functions, we obtain consistent estimates of the SFRD at z$\sim5.6$. 

As surveys of LAE at these redshifts use a sample selection based on the Ly$\alpha$ flux, we now estimate the fraction of the SFRD which is contained in the bright end of the Ly$\alpha$ luminosity function. Limiting the sample to LAEs  chosen to have EW>25\AA, commonly used in the literature \citep{ouchi_subaru/xmm-newton_2008, santos_lyman-alpha_2016} and corresponding roughly to galaxies with $\log(L_{Ly\alpha})>42.5$, we find that the SFRD from LAEs with EW>25\AA~include 75\% of the total $SFRD_{UV}$.

Estimates of the SFRD from the UV and Ly$\alpha$ luminosity functions both depend on how accurately one corrects for dust, as well as for IGM absorption for the latter.  The properties and the amount of dust in high redshift galaxies remain very uncertain and poorly constrained by current IR/submm data \citep{casey_brightest_2018}. Therefore, a better estimation of the amount of dust at z>5 is necessary. Recently, \citep{bowler_obscured_2018} discovered a galaxy with a substantial dust obscuration already at z$\sim7$. If dust plays an important role in obscuring high redshift galaxies, the total SFRD at these redshifts may then be even higher than derived from UV-selected samples. Observations of the infrared to submm continuum of these galaxies are necessary to obtain more robust estimates of the total SFRD.
While there is some indications of a low dust content in galaxies in our sample, such as those with the steepest $\beta$-slopes (see \ref{betaslopes}), it is not possible with the available data to give more robust constraints.

We also note that if reionization ended much later than z$\sim$6.6, a major fraction of Ly$\alpha$ emitting galaxies would be hidden at these redshifts. It has been previously shown that the fraction of Ly$\alpha$ emitters drops above z$\sim6$ \citep{Stark2010,Pentericci2011,Schenker2012}. This effect should be strong for our sample and contribute to substantially reduce the observed $Ly\alpha$ luminosity density. We will discuss this in detail in a forthcoming paper.

\section{Summary and conclusions}
\label{summary}

In this paper we present a sample of 52 galaxies spectroscopically confirmed at redshifts 5.0<z<6.6 and give simultaneously statistically robust constraints on the bright end of the Ly$\alpha$ and UV luminosity functions. We carefully select galaxies using several criteria including redshift verification, ensuring a high completeness and purity. This work extends the results previously obtained  to the highest redshifts probed by spectroscopic surveys \citep{cassata_vimos_2011, tasca_evolving_2015}.

We observe  galaxy number densities for the UV luminosity function somewhat higher than reported in previous works (Fig. \ref{UVLF}) but comparable to the deepest results from \cite{bouwens_uv_2015}. The main difference between our sample and previous work is the different selection technique: in previous works \citep{bowler_galaxy_2015,bouwens_uv_2015}  galaxies were selected based only on photometric properties, using the dropout technique. 

In this study, we produced a list of candidate galaxies selected from three complementary photometric techniques:  photometric redshifts, the dropout technique and the narrow band technique. These candidates are followed up with  spectroscopy to establish the redshift, and they need to satisfy a rigorous set of spectroscopic and photometric criteria  to make it in our final sample. This allows us to explore a larger parameter space and select galaxies with a broad range of properties, including galaxies with a strong UV-continuum, with or without Ly$\alpha$ in emission, but also galaxies with a less pronounced continuum break and with Ly$\alpha$ in emission.

Our main results can be summarized as follows:
\begin{itemize}

\item We observe a main sequence of  galaxies in the SFR vs. stellar mass plane, extending previous results \citep{tasca_evolving_2015} to higher redshifts z>5.0. We find no strong evidence for a turn-over of the main sequence at the massive end, indicating that star-formation quenching is not yet effective at these redshifts. We find that the normalization of the main sequence does not show any strong evolution above z$\sim3.5$. 

\item We find that the sSFR at z>5.0 remains similar as for 4.5<z<5.5. The evolution of the sSFR therefore clearly flattens at z>3 and up to z$\sim$6, at odds with current models \citep{dave_galaxy_2011, sparre_star_2015}. 

\item The brightest galaxies at z>5 are very diverse. Some have strong Ly$\alpha$ emission, others do not have Ly$\alpha$ emission at all. Some galaxies have steeper UV-continuum $\beta$-slopes than previously observed at this redshift \citep{bouwens_uv-continuum_2014}, which is observed on both spectra and photometry, while other galaxies have a flatter $\beta$-slope indicating that some dust is present. Young dust poor galaxies are mixed with older more dusty galaxies.

\item  We find that the UV luminosity function at $z\sim5.6$ can be represented by either a DPL or a Schechter function, with only a marginal preference for DPL. The UV luminosity function is comparable to other recent work \citep{bouwens_uv_2015} and the integrated UV-based SFRD is 0.27 dex higher than the median reported by \cite{madau_cosmic_2014} at the mean redshift z=5.6 of our sample.

\item We find a higher number density than previous studies on the bright end of the Ly$\alpha$ luminosity function due to our ability to find rare bright emitters thanks to the large volume probed. We find it difficult to reconcile the high number density of bright galaxies that we find with the very steep faint end slope found by the MUSE observations \citep{drake_muse_2017}, in a satisfactory fit with a Schechter function. Our results rather favor a shallower slope of the Ly$\alpha$ luminosity function of $\alpha \sim -1.7$, similar to the slope of the UV luminosity function at this redshift. Despite the large uncertainties on the faint end slope, we provide constraints to the SFRD associated to Ly$\alpha$ emitters. 

\item As we use the same sample for the UV and Ly$\alpha$ luminosity functions, we are able to compute the $SFRD_{Ly\alpha}/SFRD_{UV}$ ratio in a fully consistent way. Correcting the SFRD estimated from the Ly$\alpha$ luminosity function for IGM absorption derived from spectral modeling of the observed spectra, we obtain very similar SFRD estimates from both the UV and Ly$\alpha$ luminosity functions. Limiting our analysis to LAE with EW>25\AA, the SFRD included in these bright emitters is $\sim$75\% of the SFRD derived from the UV luminosity function, which should be taken into account when estimating the SFRD from surveys based on LAE selection. 

\item While our comparative analysis of the UV and Ly$\alpha$ SFRD favors a low dust content in most galaxies at z$\sim5.6$, measuring the total SFRD remains dependent on accurate IGM and dust absorption corrections, which may still hide some galaxies from current UV-based surveys. 
\end{itemize}

Our results, based on a sample of galaxies with confirmed spectroscopic redshifts, identify a higher number density of both UV-selected star-forming galaxies and Ly$\alpha$ emitters, particularly on the bright end. The SFRD derived from the corresponding luminosity functions are within the reported range of previous measurements, and the steep decrease of the UV SFRD above z=2 is confirmed up to z$\sim$6. The preferred shape of the Ly$\alpha$ luminosity function, on the bright end as well as on the faint end still remains to be confirmed. Future IR rest-frame  surveys e.g.  with JWST, will be necessary to make further progress.
 
\begin{acknowledgements} This work is supported by funding from the European Research Council Advanced Grant ERC--2010--AdG--268107--EARLY.  We acknowledge the support from the grants PRIN-MIUR 2015 and ASI
n.I/023/12/0 and ASI n.2018-23-HH.0.
This work is based on data products made available at the CESAM data center, Laboratoire d'Astrophysique de Marseille, France. 
 \end{acknowledgements}

\bibliographystyle{aa}
\bibliography{references}

\begin{appendix}

\section{Spectra, images, and physical parameters of the whole sample}

\begin{figure*}[h]
\centering
   \includegraphics[width=12.5cm]{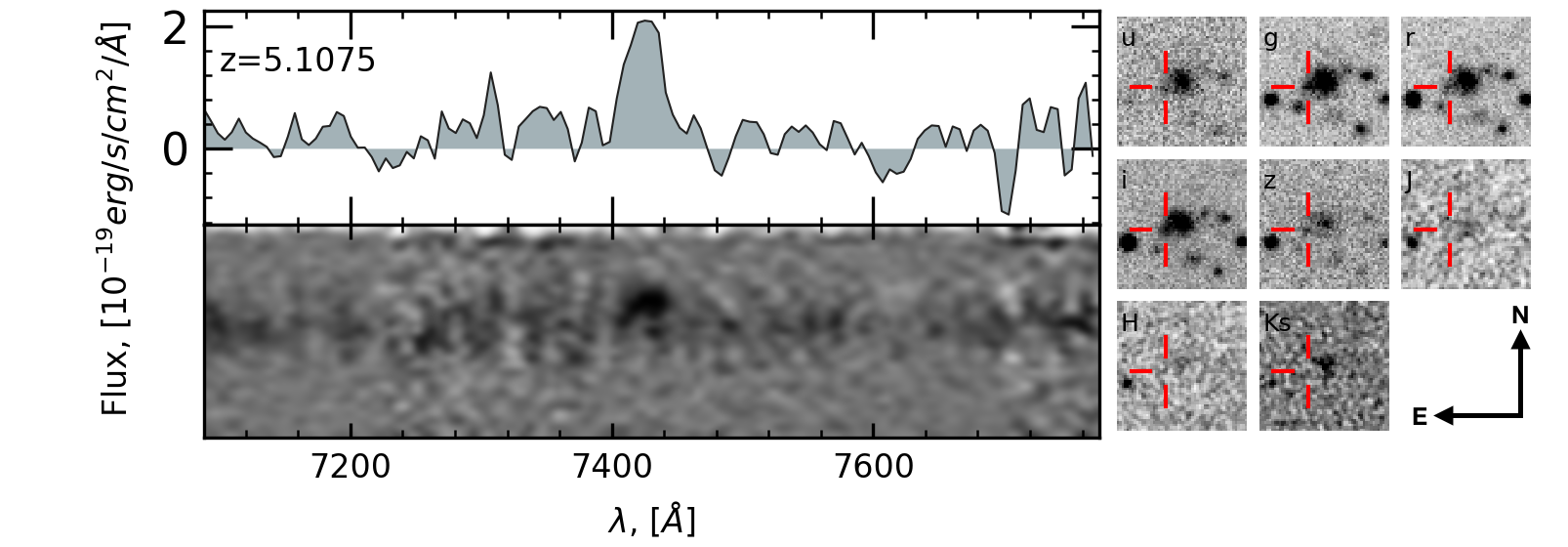}
     \caption{Spectra and 10 arcsec stamps of a galaxy with photometry contaminated by a close bright foreground object. The galaxy is pointed by red lines on the images and has an emission line seen on 1D and 2D spectra. The continuum below the emission line belongs to the bright object in the centre of the images.}
\label{contaminated}
\end{figure*}

We present the physical properties  of galaxies in our final sample in \ref{table:phys_params}, selection criteria in \ref{table:criteria}, and 1D spectra and images for individual objects in the final sample in Figures \ref{whole_sample1}. All objects are ordered by redshift. On top of 1D spectra we plot the best SED fit to photometric points, for most of the objects coinciding with the spectra. In some cases the SED templates may differ from the spectra due to slit losses. Some spectra have non-zero flux below Lyman limit 912\AA~due to either contamination by a nearby object (see Fig.\ref{contaminated}) or noise at the overlap between blue and red grism.

\begin{figure*}[h]
   \centering
   \subfloat[][]{\includegraphics[width=17cm]{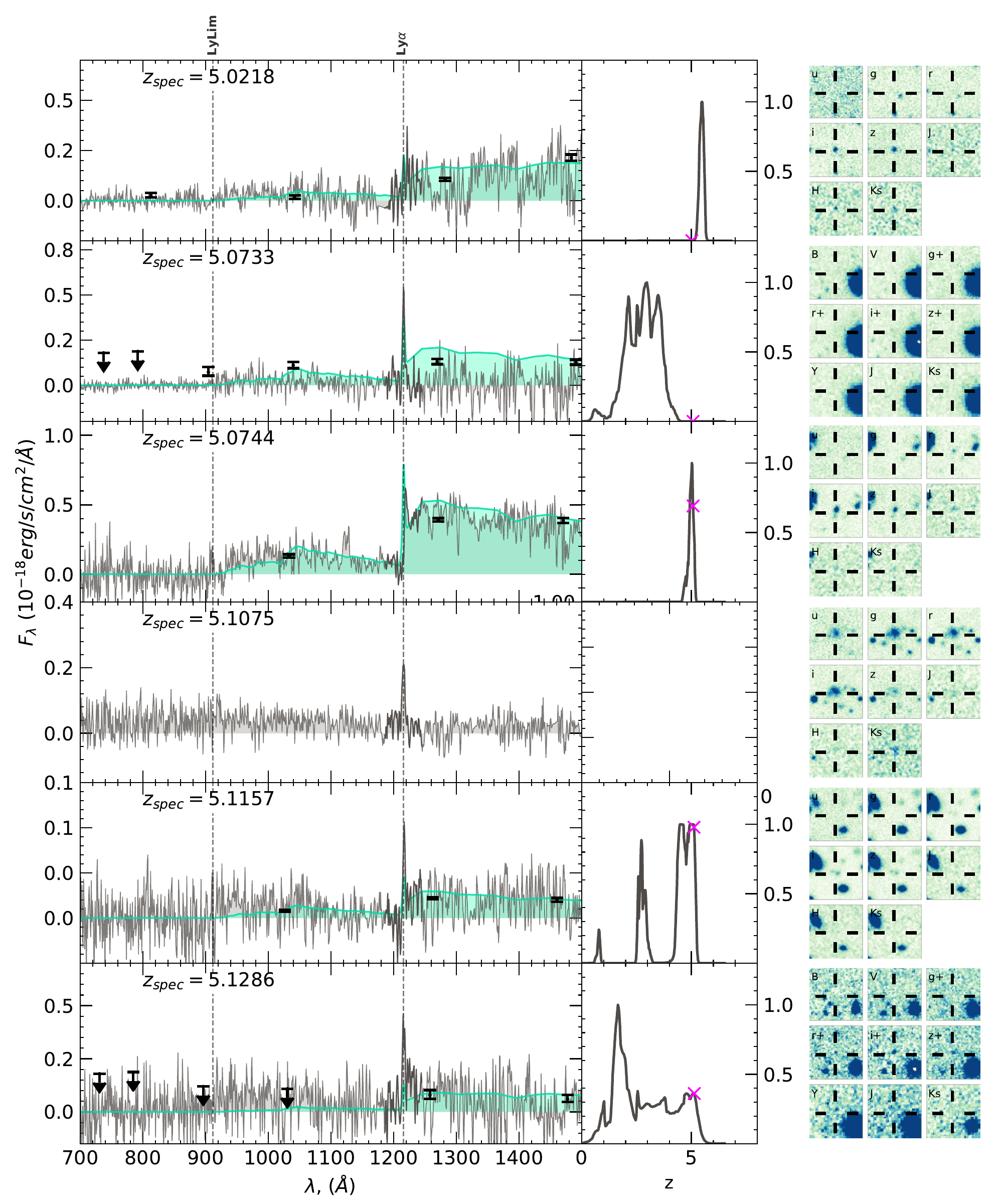}}
   \caption{Spectra and 10 arcsec image stamps of the galaxies in the sample ordered by redshift. All spectra are in the rest frame and plotted in grey. For galaxies with available photometry the photometric points are plotted in black and the best SED in light green. The PDF of photometric redshifts is plotted on the side with the real redshift marked as magenta cross.}
   \label{whole_sample1}
\end{figure*}

\begin{figure*}[h]
   \ContinuedFloat
   \centering
   \subfloat[][]{\includegraphics[width=17cm]{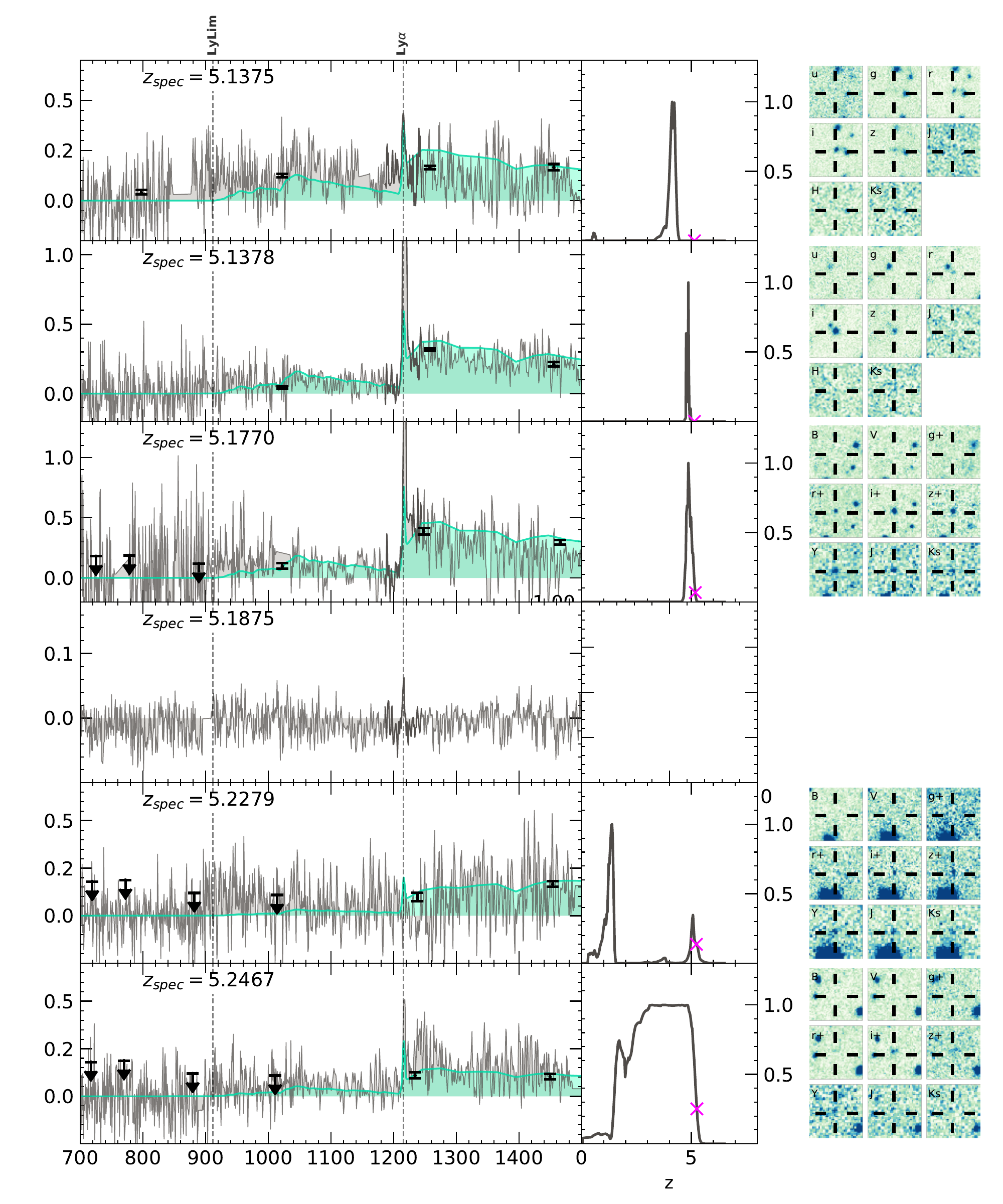}}
   \caption{Same as in Fig. \ref{whole_sample1}}
   \label{whole_sample2}
\end{figure*}

\begin{figure*}[h]
   \ContinuedFloat
   \centering
   \subfloat[][]{\includegraphics[width=17cm]{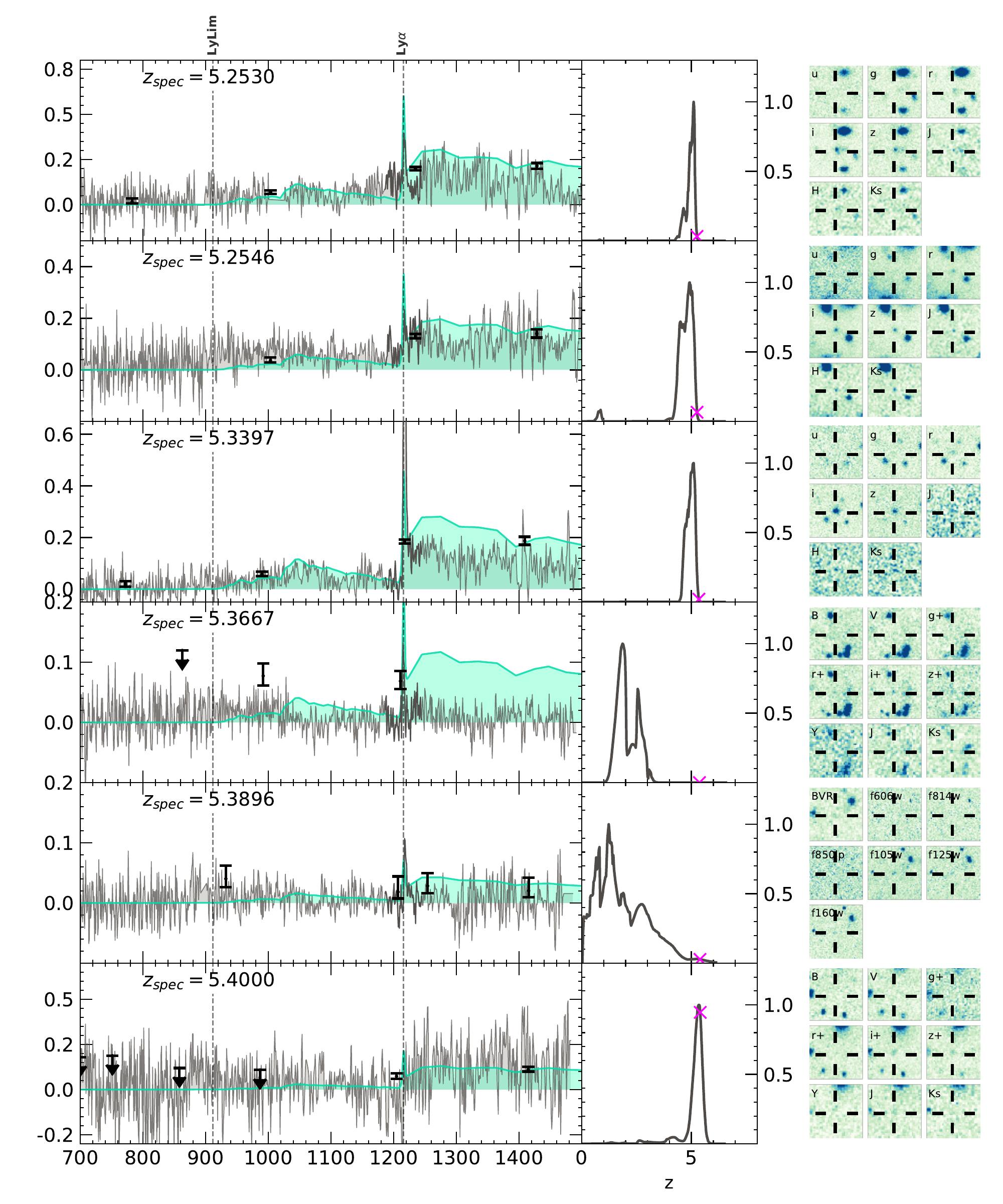}}
   \caption{Same as in Fig. \ref{whole_sample1}}
   \label{whole_sample3}
\end{figure*}

\begin{figure*}[h]
   \ContinuedFloat
   \centering
   \subfloat[][]{\includegraphics[width=17cm]{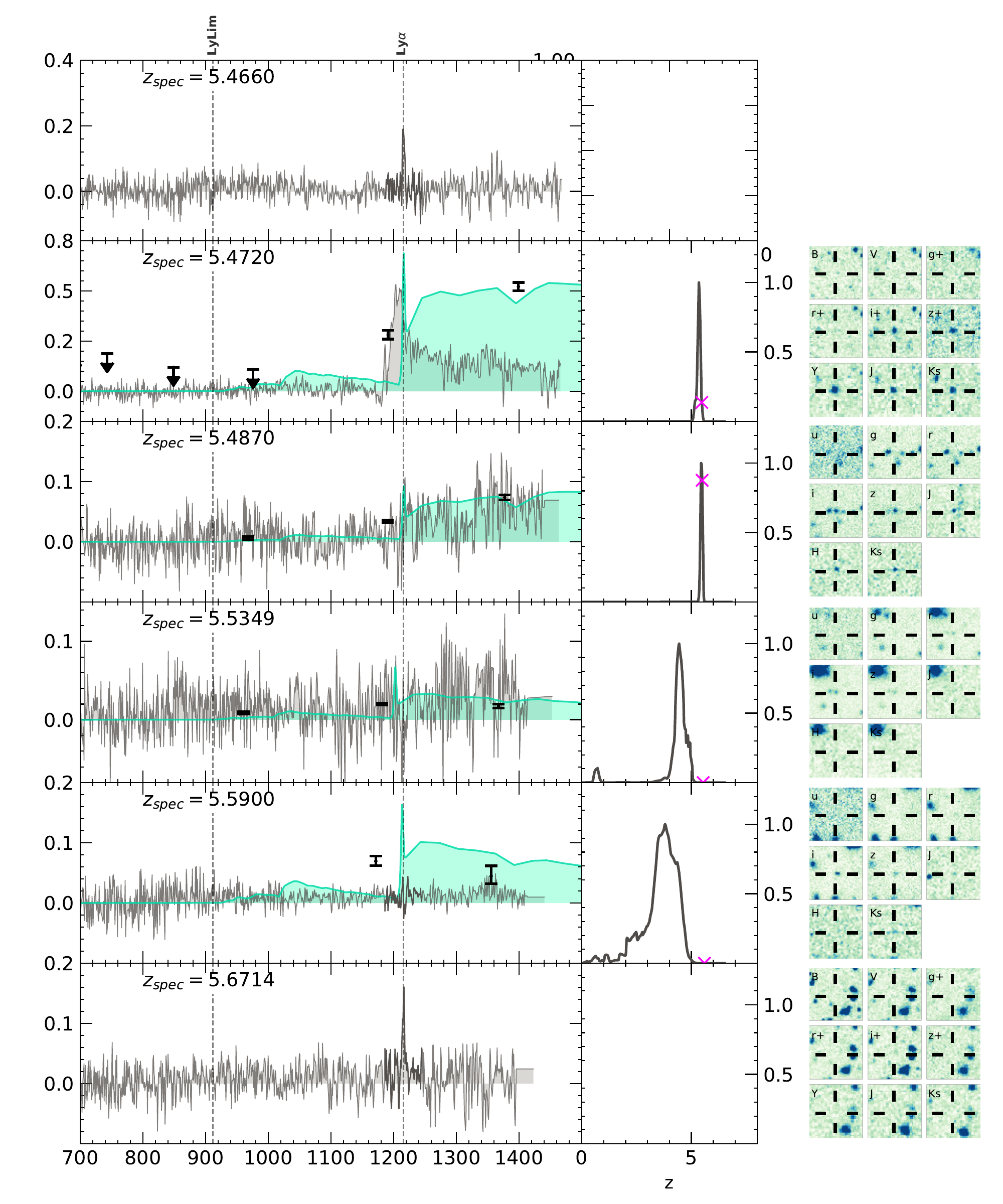}}
   \caption{Same as in Fig. \ref{whole_sample1}}
   \label{whole_sample4}
\end{figure*}

\begin{figure*}[h]
   \ContinuedFloat
   \centering
   \subfloat[][]{\includegraphics[width=17cm]{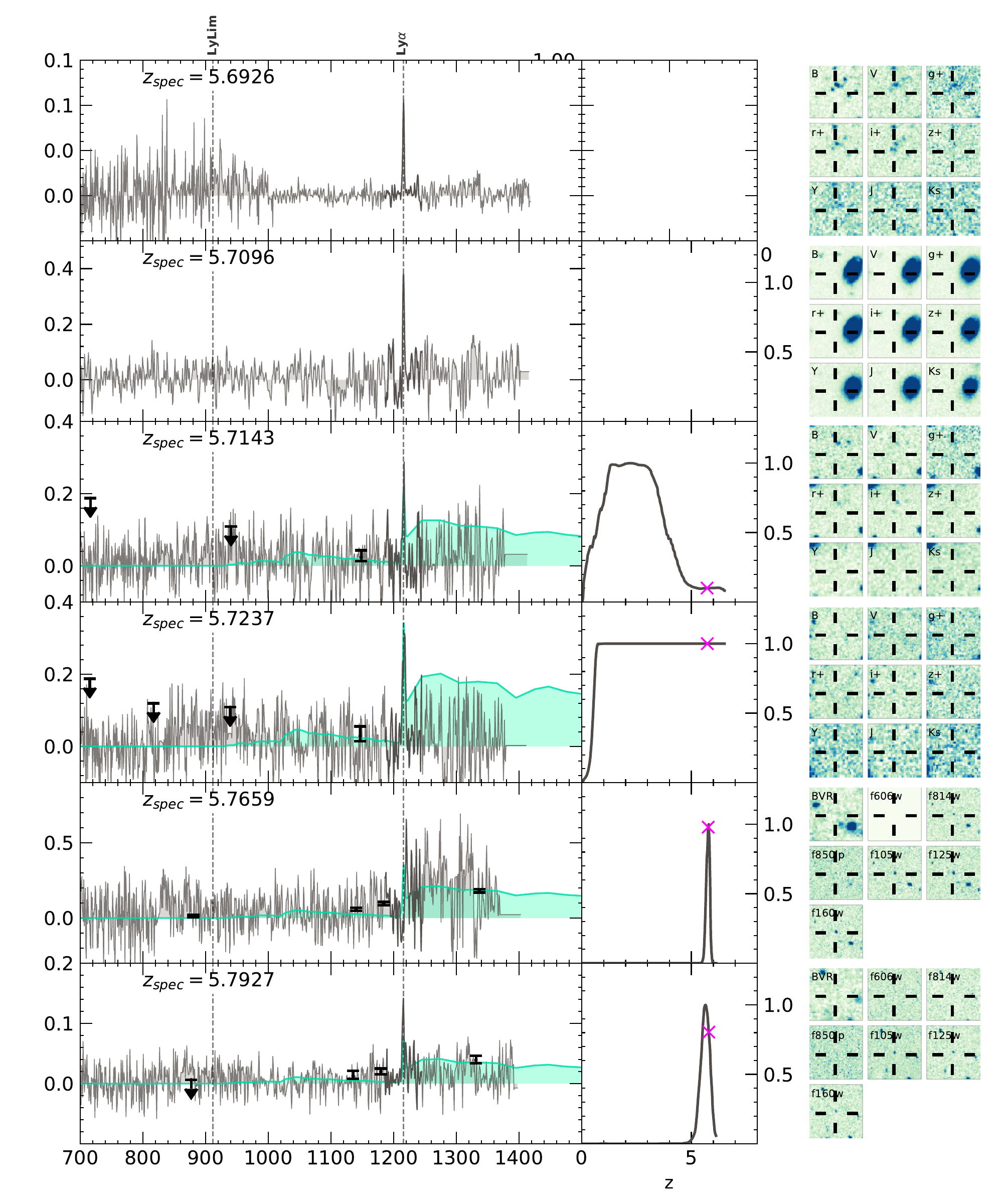}}
   \caption{Same as in Fig. \ref{whole_sample1}}
   \label{whole_sample5}
\end{figure*}

\begin{figure*}[h]
   \ContinuedFloat
   \centering
   \subfloat[][]{\includegraphics[width=17cm]{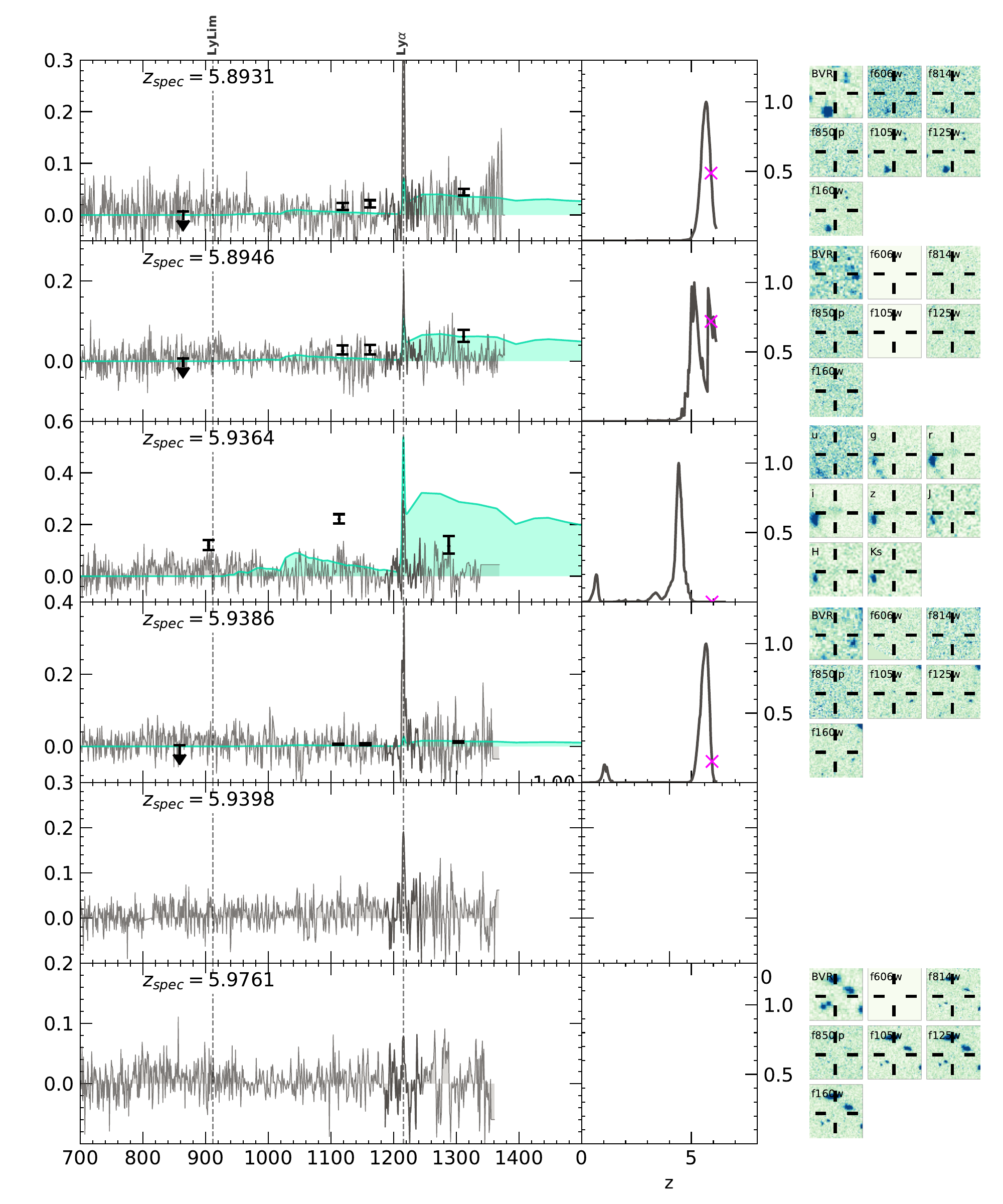}}
   \caption{Same as in Fig. \ref{whole_sample1}}
   \label{whole_sample6}
\end{figure*}

\begin{figure*}[h]
   \ContinuedFloat
   \centering
   \subfloat[][]{\includegraphics[width=17cm]{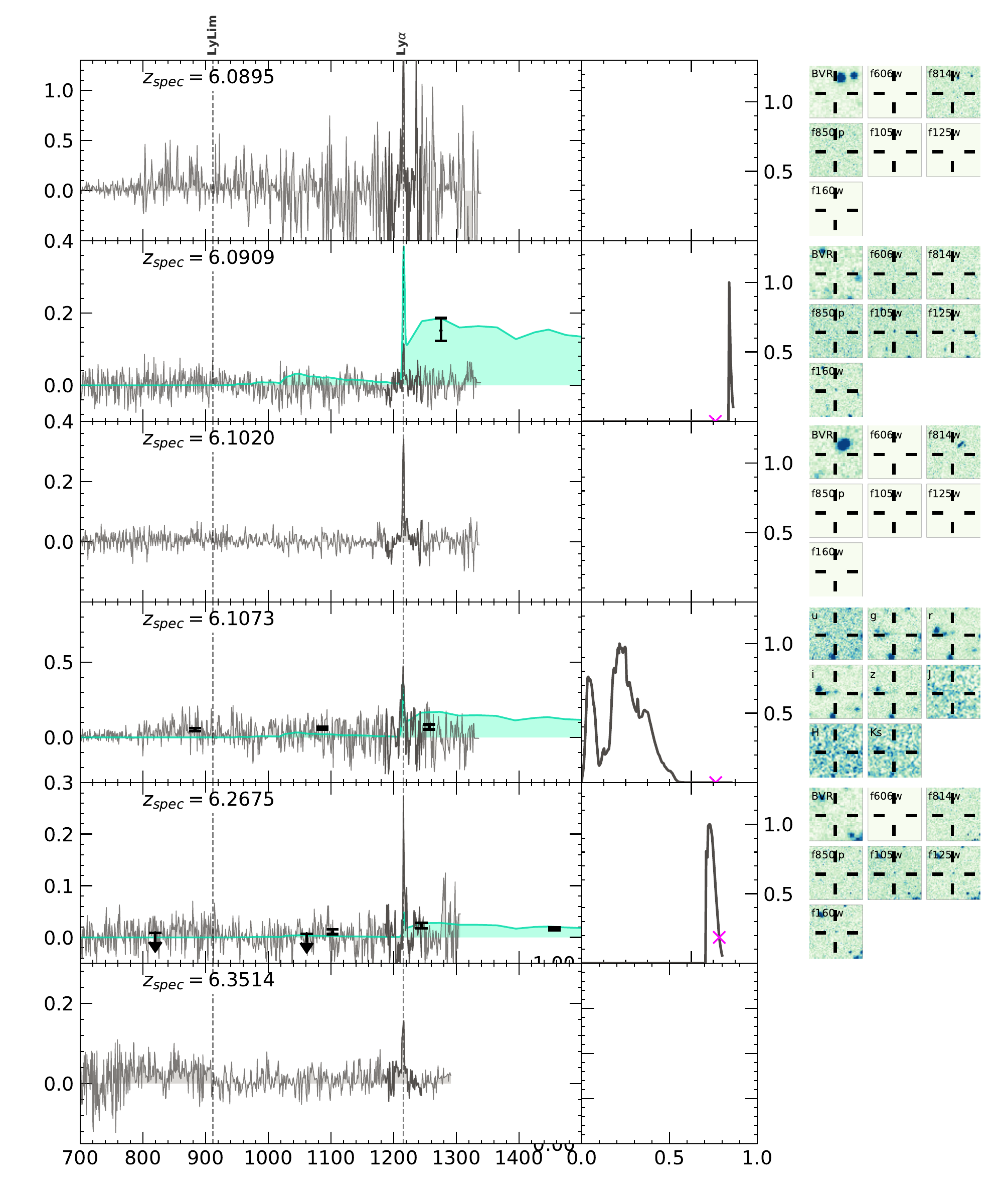}}
   \caption{Same as in Fig. \ref{whole_sample1}}
   \label{whole_sample7}
\end{figure*}

\begin{figure*}[h]
   \ContinuedFloat
   \centering
   \subfloat[][]{\includegraphics[width=17cm]{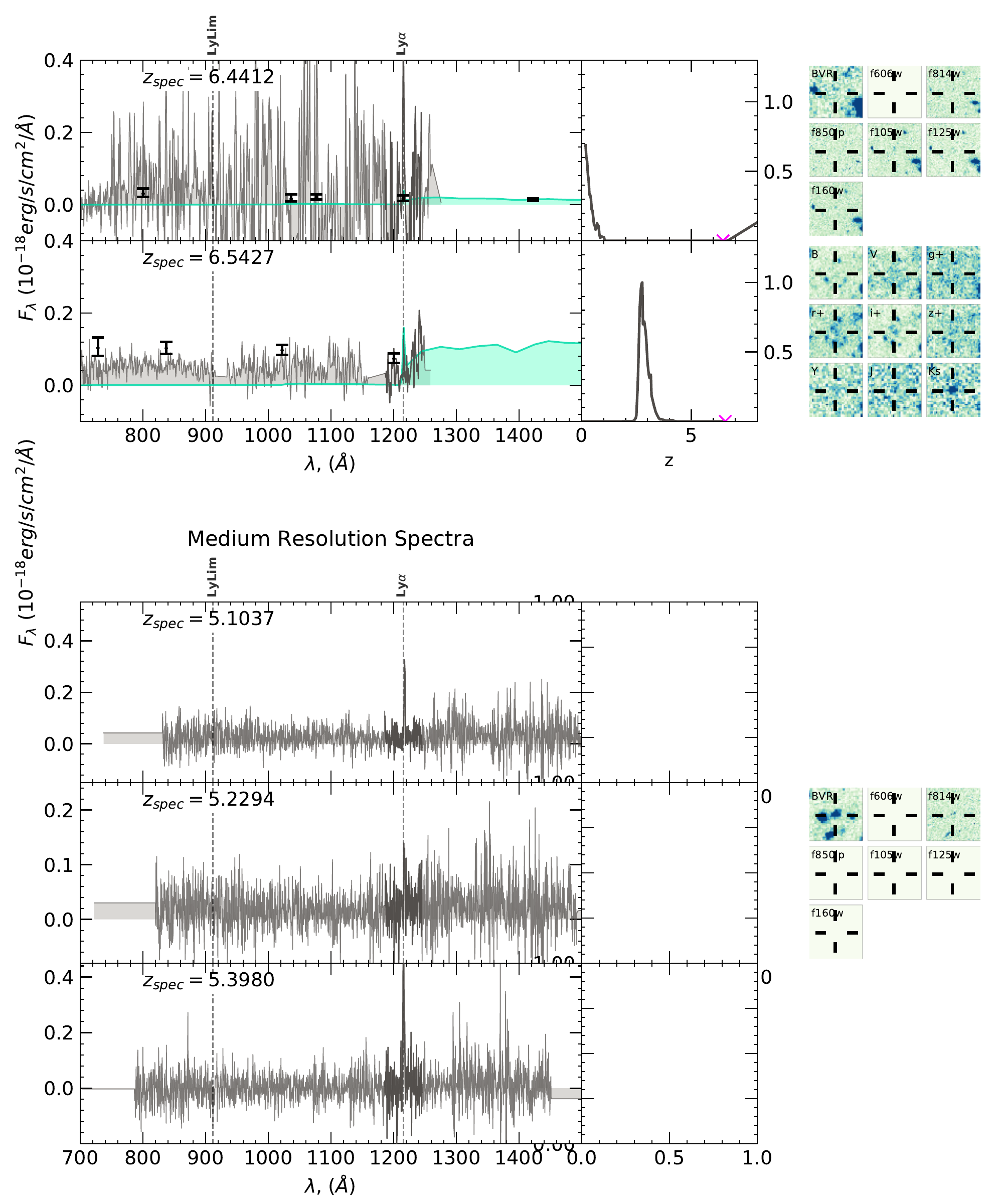}}
   \caption{Same as in Fig. \ref{whole_sample1}}
   \label{whole_sample8}
\end{figure*}

\begin{figure*}[h]
   \ContinuedFloat
   \centering
   \subfloat[][]{\includegraphics[width=17cm]{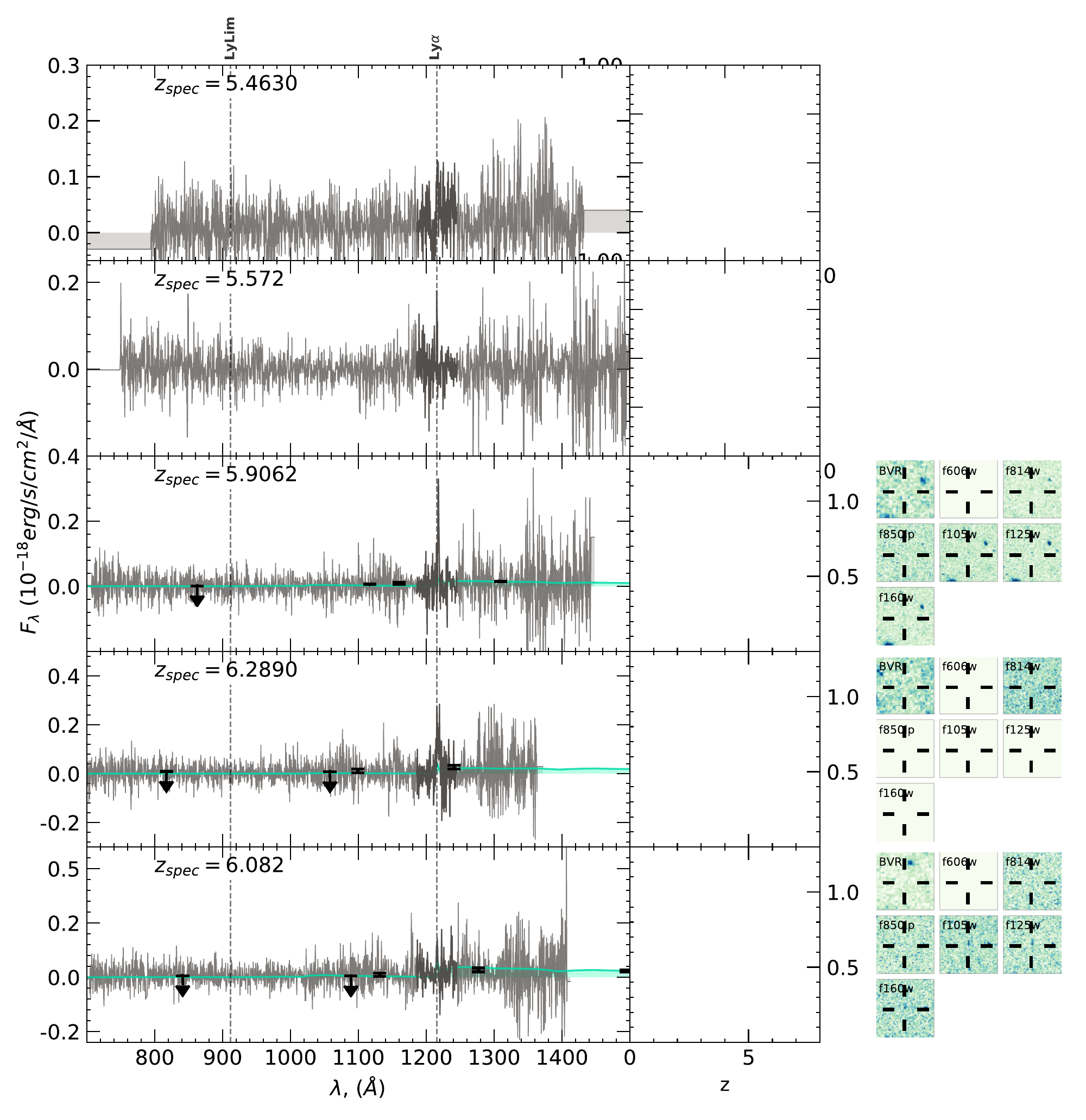}}
   \caption{Same as in Fig. \ref{whole_sample1}}
   \label{whole_sample9}
\end{figure*}

\begin{table*}
\caption{Physical parameters of galaxies in our sample} 
\label{table:phys_params}
\centering 
\begin{tabular}{c c c c c c c  c c c} 
\hline\hline 
ID & $z_{spec}$ & flag & $M_{FUV}$ & $F_{Ly\alpha}/10^{-18}$ &  SNR$_{EL}$& Age & $\log{}M_*/M_{\odot}$ & $\log{}SFR$ & $\log{}sSFR$ \\ 
 &  &  &  & ($erg/s/cm^2/\AA$) &   & (Myr) & & $(M_{\odot}yr^{-1})$ & $(Gyr^{-1})$ \\ 
\hline                               
526044493 & 5.0218 & 1 &  -22.000 &  -- & -- &   $319.4_{-194.2}^{+332.6}$ &  $ 10.3_{-  0.3}^{+  0.3}$ &  $  2.0_{-  0.1}^{+  0.2}$ &  $ -8.3_{-  0.3}^{+  0.4}$ \\ 
510352126 & 5.0733 & 9 &  -- &  $ 12.52_{- 4.29}^{+ 4.19}$ &  24.8 &  -- &  -- &  -- &  -- \\ 
520447853 & 5.0744 & 4 &  -22.563 &  $ 9.96_{- 3.94}^{+ 4.48}$ &  20.0 &  $331.8_{-200.3}^{+341.6}$ &  $ 10.1_{-  0.3}^{+  0.2}$ &  $  1.8_{-  0.2}^{+  0.2}$ &  $ -8.3_{-  0.3}^{+  0.4}$ \\ 
520326980 & 5.1075 & 39 &  -22.932 &  $ 5.99_{- 1.39}^{+ 1.41}$ &  27.2 &  -- &  -- &  -- &  -- \\ 
520091023 & 5.1157 & 29 &  -22.409 &  $ 2.05_{- 0.32}^{+ 0.56}$ &  16.5 &  $329.3_{-216.4}^{+411.7}$ &  $  9.9_{-  0.4}^{+  0.4}$ &  $  1.6_{-  0.2}^{+  0.2}$ &  $ -8.3_{-  0.4}^{+  0.5}$ \\ 
5181313821 & 5.1286 & 3 &  -20.665 &  $ 10.19_{- 6.78}^{+ -8.48}$ &  17.4 &  $386.5_{-242.3}^{+393.3}$ &  $ 10.1_{-  0.3}^{+  0.2}$ &  $  1.7_{-  0.3}^{+  0.4}$ &  $ -8.4_{-  0.3}^{+  0.4}$ \\ 
526098666 & 5.1375 & 2 &  -21.653 &  $ 8.44_{- 3.81}^{+ 4.32}$ &  13.7 &  $156.5_{-106.0}^{+528.9}$ &  $  9.1_{-  0.4}^{+  0.7}$ &  $  1.2_{-  0.1}^{+  0.1}$ &  $ -7.9_{-  0.7}^{+  0.6}$ \\ 
520180097 & 5.1378 & 3 &  -21.931 &  $ 119.20_{- 3.44}^{+ 7.80}$ &  216.7 &  $152.7_{- 99.0}^{+245.2}$ &  $  9.6_{-  0.3}^{+  0.3}$ &  $  1.6_{-  0.2}^{+  0.1}$ &  $ -8.0_{-  0.4}^{+  0.4}$ \\ 
5100816509 & 5.1770 & 4 &  -22.320 &  $ 58.08_{- 13.11}^{+ 9.19}$ &  53.0 &  $278.6_{-170.8}^{+321.8}$ &  $  9.8_{-  0.3}^{+  0.3}$ &  $  1.5_{-  0.1}^{+  0.2}$ &  $ -8.2_{-  0.4}^{+  0.4}$ \\ 
F51P008\_23 & 5.1875 & 29 &  -- &  $ 0.96_{- 0.06}^{+ 0.40}$ &  8.7 &  -- &  -- &  -- &  -- \\ 
5180845047 & 5.2279 & 2 &  -21.903 &  -- & -- &   $ 53.0_{-  5.8}^{+ 67.2}$ &  $  9.9_{-  0.1}^{+  0.1}$ &  $  2.3_{-  0.1}^{+  0.1}$ &  $ -7.6_{-  0.2}^{+  0.3}$ \\ 
5181204998 & 5.2467 & 2 &  -21.245 &  $ 9.71_{- 3.04}^{+ 2.64}$  & 15.4 &  $412.9_{-265.3}^{+402.6}$ &  $ 10.4_{-  0.7}^{+  0.6}$ &  $  1.8_{-  0.5}^{+  0.6}$ &  $ -8.4_{-  0.4}^{+  0.4}$ \\ 
527022086 & 5.2530 & 2 &  -21.983 &  $ 6.07_{- 2.78}^{+ 3.05}$ &  11.0 &  $439.2_{-288.4}^{+392.4}$ &  $ 10.0_{-  0.4}^{+  0.4}$ &  $  1.5_{-  0.2}^{+  0.2}$ &  $ -8.4_{-  0.3}^{+  0.5}$ \\ 
526008600 & 5.2546 & 2 &  -21.560 &  -- & -- &   $480.8_{-311.3}^{+386.6}$ &  $ 10.4_{-  0.5}^{+  0.4}$ &  $  1.9_{-  0.3}^{+  0.5}$ &  $ -8.5_{-  0.4}^{+  0.5}$ \\ 
520348474 & 5.3397 & 4 &  -21.803 &  $ 37.11_{- 1.53}^{+ 3.44}$ &  95.5 &  $292.5_{-191.6}^{+396.8}$ &  $  9.6_{-  0.4}^{+  0.4}$ &  $  1.3_{-  0.1}^{+  0.2}$ &  $ -8.2_{-  0.4}^{+  0.5}$ \\ 
5101263627 & 5.3667 & 1 &  -21.153 &  $ 2.39_{- 0.56}^{+ 0.64}$ &  17.1 &  $625.2_{-331.7}^{+292.1}$ &  $ 10.4_{-  0.2}^{+  0.2}$ &  $  1.8_{-  0.3}^{+  0.4}$ &  $ -8.7_{-  0.4}^{+  0.4}$ \\ 
532000128 & 5.3896 & 1 &  -20.051 &  $ 2.09_{- 0.28}^{+ 0.72}$ &  21.5 &  $306.8_{-191.2}^{+345.8}$ &  $  9.0_{-  0.4}^{+  0.3}$ &  $  0.7_{-  0.2}^{+  0.3}$ &  $ -8.3_{-  0.4}^{+  0.4}$ \\ 
5180752864 & 5.4000 & 2 &  -21.425 &  -- & -- &   $477.4_{-256.2}^{+358.9}$ &  $ 10.7_{-  0.2}^{+  0.2}$ &  $  2.2_{-  0.5}^{+  0.1}$ &  $ -8.5_{-  0.5}^{+  0.4}$ \\ 
F52P002\_135 & 5.4660 & 29 &  -- &  $ 4.68_{- 0.35}^{+ 1.11}$  & 20.2 &  -- &  -- &  -- &  -- \\ 
5101448618 & 5.4720 & 14 &  -23.208 &  $ 61.98_{- 2.87}^{+ 8.56}$ &  68.8 &  $159.5_{-104.5}^{+194.8}$ &  $ 10.5_{-  0.2}^{+  0.2}$ &  $  2.5_{-  0.2}^{+  0.1}$ &  $ -7.9_{-  0.5}^{+  0.3}$ \\ 
528295041 & 5.4870 & 1 &  -22.451 &  -- & -- &   $131.5_{- 80.1}^{+155.5}$ &  $ 10.2_{-  0.2}^{+  0.2}$ &  $  2.3_{-  0.1}^{+  0.3}$ &  $ -8.0_{-  0.2}^{+  0.4}$ \\ 
528471411 & 5.5349 & 1 &  -21.447 &  -- & -- &  $365.4_{-256.9}^{+436.2}$ &  $  9.7_{-  0.5}^{+  0.4}$ &  $  1.3_{-  0.1}^{+  0.2}$ &  $ -8.3_{-  0.4}^{+  0.6}$ \\ 
526136058 & 5.5900 & 1 &  -20.272 &  -- & -- &  $354.8_{-236.8}^{+399.9}$ &  $  9.4_{-  0.5}^{+  0.4}$ &  $  1.1_{-  0.2}^{+  0.3}$ &  $ -8.3_{-  0.4}^{+  0.5}$ \\ 
519816038 & 5.6714 & 9 &  -- &  $ 4.55_{- 1.51}^{+ 1.80}$ &  21.0 &  -- &  -- &  -- &  -- \\
5150100073 & 5.6926 & 9 &  -- &  $ 2.06_{- 0.37}^{+ 0.31}$ &  35.4 &  -- &  -- &  -- &  -- \\ 
519816019 & 5.7096 & 9 &  -- &  $ 1.37_{- 0.95}^{+ -1.14}$ &  10.5 &  -- &  -- &  -- &  -- \\ 
5150100059 & 5.7143 & 9 &  -21.301 &  $ 5.23_{- 2.37}^{+ 2.52}$ &  12.2 &  $333.0_{-207.2}^{+334.2}$ &  $  9.5_{-  0.3}^{+  0.3}$ &  $  1.2_{-  0.2}^{+  0.3}$ &  $ -8.3_{-  0.4}^{+  0.4}$ \\ 
5150100005 & 5.7237 & 9 &  -21.939 &  $ 10.07_{- 0.47}^{+ 4.14}$ &  17.8 &  $363.8_{-223.8}^{+336.2}$ &  $ 10.2_{-  0.3}^{+  0.3}$ &  $  1.8_{-  0.3}^{+  0.4}$ &  $ -8.4_{-  0.3}^{+  0.4}$ \\ 
532000001 & 5.7659 & 2 &  -21.951 &  -- & -- &  $421.6_{-226.0}^{+298.2}$ &  $  9.9_{-  0.2}^{+  0.1}$ &  $  1.5_{-  0.2}^{+  0.2}$ &  $ -8.4_{-  0.3}^{+  0.3}$ \\ 
532000014 & 5.7927 & 9 &  -20.136 &  $ 2.59_{- 0.27}^{+ 0.49}$ &  15.5 &  $355.1_{-222.8}^{+335.0}$ &  $  9.1_{-  0.4}^{+  0.3}$ &  $  0.8_{-  0.2}^{+  0.2}$ &  $ -8.3_{-  0.3}^{+  0.4}$ \\ 
532000022 & 5.8931 & 9 &  -20.118 &  $ 10.62_{- 0.22}^{+ 1.03}$ &  38.7 &  $316.8_{-203.4}^{+340.0}$ &  $  8.9_{-  0.4}^{+  0.3}$ &  $  0.6_{-  0.1}^{+  0.2}$ &  $ -8.3_{-  0.4}^{+  0.4}$ \\ 
532000085 & 5.8946 & 9 &  -20.803 &  $ 3.50_{- 0.15}^{+ 0.81}$ &  16.9 &  $167.2_{-112.0}^{+242.6}$ &  $  9.0_{-  0.3}^{+  0.3}$ &  $  1.0_{-  0.2}^{+  0.2}$ &  $ -8.0_{-  0.4}^{+  0.4}$ \\ 
520262147 & 5.9364 & 1 &  -- &  $ 7.48_{- 0.68}^{+ 1.25}$ &  17.6 &  -- &  -- &  -- &  -- \\ 
532000123 & 5.9386 & 9 &  -19.274 &  $ 10.77_{- 0.79}^{+ 1.57}$ &  30.6 &  $381.3_{-244.0}^{+324.1}$ &  $  9.1_{-  0.5}^{+  0.8}$ &  $  0.7_{-  0.4}^{+  0.7}$ &  $ -8.4_{-  0.3}^{+  0.4}$ \\ 
F52P02B\_26 & 5.9398 & 29 &  -- &  $ 4.82_{- 0.78}^{+ 1.61}$ &  14.6 &  -- &  -- &  -- &  -- \\ 
532000341 & 5.9761 & 29 &  -- &  $ 1.84_{- 0.46}^{+ 0.92}$ &  6.4 & -- &  -- &  -- &  -- \\ 
532000150 & 6.0895 & 29 &  -- &  $ 26.58_{- 0.63}^{+ 8.83}$ & 6.8 &  -- &  -- &  -- &  -- \\ 
534024726 & 6.0909 & 9 	&  -- &  $ 1.73_{- 0.19}^{+ 0.62}$ &  7.9 &  -- &  -- &  -- &  -- \\ 
532000024 & 6.1020 & 9 &  -- &  $ 6.70_{- 0.41}^{+ 0.71}$ &  26.9 &  -- &  -- &  -- &  -- \\ 
520460800 & 6.1073 & 9 &  -21.429 &  $ 13.87_{- 1.71}^{+ 9.94}$ &  15.9 &  $355.2_{-223.8}^{+316.8}$ &  $ 10.0_{-  0.4}^{+  0.4}$ &  $  1.7_{-  0.3}^{+  0.3}$ &  $ -8.4_{-  0.3}^{+  0.4}$ \\ 
532000053 & 6.2675 & 29 &  -19.839 &  $ 3.13_{- 0.13}^{+ 0.52}$ &  11.9 &  $272.7_{-172.3}^{+302.6}$ &  $  9.3_{-  0.7}^{+  0.8}$ &  $  1.0_{-  0.4}^{+  0.7}$ &  $ -8.2_{-  0.4}^{+  0.5}$ \\ 
F52P004\_32 & 6.3514 & 29 &  -- &  $ 5.36_{- 0.83}^{+ 1.51}$ &  40.6 &  -- &  -- &  -- &  -- \\ 
534066387 & 6.4412 & 29 &  -19.658 &  $ 6.15_{- 1.04}^{+ 3.30}$  & 6.9 &  $309.1_{-192.4}^{+292.3}$ &  $  9.0_{-  0.4}^{+  0.4}$ &  $  0.7_{-  0.2}^{+  0.4}$ &  $ -8.3_{-  0.4}^{+  0.4}$ \\ 
5170041461 & 6.5427 & 9 &  -- &  $ 1.00_{- -0.34}^{+ 0.74}$ &  2.9 &  -- &  -- &  -- &  -- \\ 
\hline  
F53P003\_32 & 5.1037 & 9 &  -- &  $ 3.67_{- 0.24}^{+ 0.56}$  & 16.8 &  -- &  -- &  -- &  -- \\ 
530045549 & 5.2294 & 9 &  -- &  $ 1.24_{- 1.95}^{+ 0.57}$ &  5.0 &  -- &  -- &  -- &  -- \\ 
F53P003\_Q1\_20 & 5.3980 & 4 &  -- &  $ 6.60_{- 0.78}^{+ 1.72}$ & 17.6 &  -- &  -- &  -- &  -- \\
F53P003\_Q3\_34 & 5.4630 & 2 &  -- &  $ 1.10_{- 0.68}^{+ 1.40}$ &  4.0 &  -- &  -- &  -- &  -- \\ 
F53P003\_Q2\_26 & 5.5720 & 1 &  -- &  $1.30_{- 0.00}^{+ 0.54}$ &  5.8 &  -- &  -- &  -- &  -- \\ 
532000189 & 5.9062 & 9 &  -- &  $ 3.57_{- 0.39}^{+ 0.56}$ &  12.9 &  $343.4_{-213.7}^{+333.1}$ &  $  9.3_{-  0.6}^{+  0.7}$ &  $  0.8_{-  0.4}^{+  0.7}$ &  $ -8.5_{-  0.3}^{+  0.5}$ \\ 
532000086 & 6.2890 & 9 &  -- &  $ 5.74_{- 0.91}^{+ 2.05}$ &  13.8 &  $420.8_{-255.6}^{+278.9}$ &  $  10.1_{-  0.2}^{+  0.1}$ &  $  1.7_{-  0.4}^{+  0.2}$ &  $ -8.4_{-  0.4}^{+  0.4}$ \\ 
532000101 & 6.0820 & 9 &  -- &  $ 1.94_{- 0.38}^{+ 0.98}$ & 3.6 &  $343.7_{-214.0}^{+310.8}$ &  $  9.7_{-  0.6}^{+  0.4}$ &  $  1.4_{-  0.5}^{+  0.4}$ &  $ -8.3_{-  0.3}^{+  0.4}$ \\  
\hline  
\end{tabular}
\end{table*}

\begin{table*}
\caption{Selection criteria} 
\label{table:criteria}
\centering 
\begin{tabular}{c c c c c c c c c c  p{5cm}} 
\hline\hline 
N & ID & $z_{spec}$ & flag  & \tablefootmark{a} & \tablefootmark{b} & \tablefootmark{c} & $dz/\sigma$ & $\chi^2$ & \tablefootmark{d} & comments \\ 
\hline                               
1 & 526044493 & 5.0218 & 2   & no & yes & yes & 0.58 & 16.75 & no & faint detection in g-band \\ 
2 & 510352126 & 5.0733 &  9   & yes & yes & yes & 2.17 & 31.96 & yes  & bright object nearby \\ 
3 & 520447853 & 5.0744 &  4  & yes & yes & yes & 0.48 & 5.48 & no & \\ 
4 & 520326980 & 5.1075 &  39  & no & yes & yes & - & - & yes & contaminated photometry \\ 
5 & 520091023 & 5.1157 &  29  & yes & yes & yes & 0.31 & 0.00 & yes & two close components \\ 
6 & 5181313821 & 5.1286 &  3  & yes & yes & yes & 0.32 & 3.13 & yes & bright object nearby \\ 
7 & 526098666 & 5.1375  & 2   & no & yes & yes & 6.45 & 57.61 & no & faint detection in g-band\\ 
8 & 520180097 & 5.1378  & 3   & yes & yes & yes & 1.28 & 53.57 & no & \\ 
9 & 5100816509 & 5.1770 & 4 	  & yes & yes & yes & 1.69 & 6.37 & no & \\ 
10 &F51P008\_23 & 5.1875 & 29  & - & - & - & - & - & no &  \\ 
11 & 5180845047 & 5.2279  & 2   & yes & yes & yes & 0.22 & 11.14 & yes & bright object nearby \\ 
12 & 5181204998 & 5.2467 & 2   & yes & yes & yes & 1.24 & 1.24 & no & detected only in 2 bands \\ 
13 & 527022086 & 5.2530 &  2   & no & yes & yes & 0.79 & 3.20 & no & faint detection in g-band \\ 
14 & 526008600 & 5.2546 &  2   & yes & yes & yes & 1.05 & 2.12 & no & \\ 
15 & 520348474 & 5.3397 &  4   & no & yes & yes & 1.20 & 5.34 & no & faint detection in g-band \\ 
16 & 5101263627 & 5.3667 & 1   & yes & no & yes & 1.82 & 12.45 & no & \\ 
17 & 532000128 & 5.3896 & 1   & yes & no & yes & 0.02 & 6.16 & yes  & two close components \\ 
18 & 5180752864 & 5.4000 &  2   & yes & yes & yes & 0.24 & 0.14 & no & \\ 
19 &F52P002\_135 & 5.4660 & 29  & - & - & - & - & - & no &  \\ 
20 & 5101448618 & 5.4720  & 14   & yes & yes & yes & 0.81 & 5.54 & no & AGN \\ 
21 & 528295041 & 5.4870 & 1   & yes & yes & yes & 2.07 & 4.88 & yes & artifacts on J image \\ 
22 & 528471411 & 5.5349 & 1   & yes & yes & yes & 3.19 & 36.76 & no & \\ 
23 & 526136058 & 5.5900 & 1  & yes & no & yes & 2.57 & 15.57 & no &  \\ 
24 & 519816038 & 5.6714 &  9   & yes & yes & yes & - & - & no & \\ 
25 & 5150100073 & 5.6926 &  9   & no & no & yes & 24.14 & 114.19 & yes & contaminated photometry \\ 
26 & 519816019 & 5.7096  & 9   & yes & yes & yes & - & - & yes & bright object nearby \\ 
27 & 5150100059 & 5.7143  & 9 & yes & yes & yes & 2.10 & 4.75 & no &  \\ 
28 & 5150100005 & 5.7237 &  9 & yes & yes & yes & 0.34 & 0.00 & no &  \\ 
29 & 532000001 & 5.7659 &  2   & yes & yes & yes & 0.30 & 3.46 & no & \\ 
30 & 532000014 & 5.7927  & 9   & yes & yes & yes & 0.44 & 1.66 & no & \\ 
31 & 532000022 & 5.8931 &  9   & yes & yes & yes & 0.35 & 5.17 & no &  \\ 
32 & 532000085 & 5.8946  & 9   & yes & yes & yes & 0.28 & 4.37 & no & \\ 
33 & 520262147 & 5.9364 & 1   & yes & no & yes & 6.30 & 117.26 & yes & unreliable photometry \\ 
34 & 532000123 & 5.9386 &  9   & yes & yes & yes & 1.33 & 2.41 & no & \\ 
35 &F52P02B\_26 & 5.9398 & 29  & - & - & - & - & - & no &  \\  
36 & 532000341 & 5.9761  & 29   & - & - & - & - & - & no & \\ 
37 & 532000150 & 6.0895  & 29   & - & - & - & - & - & no & \\ 
38 & 534024726 & 6.0909 &  9   & yes & yes & yes & 1.04 & 2.05 & no & \\ 
39 & 532000024 & 6.1020 &  9   & - & - & - & - & - & no & \\ 
40 & 520460800 & 6.1073 &  9   & no & yes & yes & 1.66 & 58.39 & no & \\ 
41 & 532000053 & 6.2675  & 29   & yes & yes & yes & 1.61 & 2.62 & no & \\ 
42 &F52P004\_32 & 6.3514 & 29  & - & - & - & - & - & no &  \\  
43 & 534066387 & 6.4412 &  29   & no & no & yes & 1.59 & 20.87 & no & \\ 
44 & 5170041461 & 6.5427 &  9   & no & no & yes & 17.36 & 134.22 & yes & contaminated photometry \\ 
\hline  
45 &F53P003\_32 & 5.1037 &  9   & - & - & - & - & - & no &  \\ 
46 & 530045549 & 5.2294 &  9   & yes & yes & yes & - & - & no &  \\ 
47 & F53P003\_20 & 5.3980 &  4   & - & - & - & - & - & no &  \\ 
48 & F53P003\_34 & 5.4630 &  2   & - & - & - & - & - & no &  \\ 
49 & F53P003\_26 & 5.572 &  9   & - & - & - & - & - & no &  \\ 
50 & 532000189 & 5.9062 &  9  & yes & yes & yes & - & 5.49 & no &  \\ 
51 & 532000086 & 6.2890 &  9  & yes & yes & yes & - & 11.64 & no &  \\ 
52 & 532000101 & 6.0820 &  9   & yes & yes & yes & - & 0.57 & no &  \\ 
\hline  
\end{tabular}
\tablefoot{
\tablefoottext{1}{No detection below 912\AA}
\tablefoottext{2}{No or faint detection between 912\AA~and Ly$\alpha$}
\tablefoottext{3}{At least one detection at position of Ly$\alpha$ or after Ly$\alpha$}
\tablefoottext{4}{Signs of contamination in photometry or bad sky subtraction }
}
\end{table*}

\end{appendix}

\end{document}